\pdfoutput=1

\documentclass{vldb}
\usepackage{booktabs}
\usepackage{graphicx}
\usepackage{epstopdf}
\usepackage{multirow}
\usepackage{amsmath,amssymb}
\usepackage[lined,boxed,vlined,ruled,linesnumbered]{algorithm2e}
\usepackage{caption}
\usepackage{tikz}
\usepackage{pgflibraryarrows}
\usepackage{pgflibrarysnakes}
\usepackage{subfigure}
\usepackage[utf8]{inputenc}
\usepackage[english]{babel}
\usepackage{hyperref}
\usepackage{dsfont}
\usepackage{cite}
\usepackage{url}
\usepackage{makecell}
\usepackage{mathrsfs}
\usepackage{booktabs}
\usepackage{url}
\usepackage{balance}
\newtheorem{definition}{Definition}

\newcommand{\stitle}[1]{\vspace{1ex}\noindent\textup{\textbf{#1.}}}

\newcommand{\indfunc}[1]{\mathds{1}_{\{#1\}}\xspace}

\makeatletter
  \newcommand\figcaption{\def\@captype{figure}\caption}
  \newcommand\tabcaption{\def\@captype{table}\caption}
\makeatother

\begin{document}
%\title{R-Crowd: Crowdsourcing Relational Data}
\title{T-Crowd: Effective Crowdsourcing for Tabular Data}

%\title{StructCrowd: A Unified Crowdsourcing Framework for Collecting Multi-type Structured Data}

\numberofauthors{1}
\author{
\alignauthor
Caihua Shan{$\,^\dag$}~~~~~~~Nikos Mamoulis{$\,^\dag$}~~Guoliang Li{$\,^\#$}\\~~~Reynold Cheng{$\,^\dag$}~~~Zhipeng Huang{$\,^\dag$}~~~Yudian Zheng{$\,^\dag$}
\\ \vspace{.3em} \affaddr{$^\dag$ Department of Computer Science, The University of Hong Kong} 
\\ \vspace{.3em} \affaddr{$^\#$Department of Computer Science, Tsinghua University} 
 \\ \email{\{chshan, nikos, ckcheng, zphuang, ydzheng2\}@cs.hku.hk,\\liguoliang@tsinghua.edu.cn}
%\vspace{-.5em}
}

% \author{Caihua Shan}
% \affiliation{
%  \institution{The University of Hong Kong}
%  }
% \email{chshan@cs.hku.hk}

% \author{Nikos Mamoulis}
% \affiliation{
%  \institution{The University of Hong Kong}
%  }
% \email{nikos@cs.hku.hk}

% \author{Guoliang Li}
% \affiliation{
%  \institution{Tsinghua University}
%  }
% \email{liguoliang@tsinghua.edu.cn}

% \author{Reynold Cheng}
% \affiliation{
%  \institution{The University of Hong Kong}
%  }
% \email{ckcheng@cs.hku.hk}

% \author{Zhipeng Huang}
% \affiliation{
%  \institution{The University of Hong Kong}
%  }
% \email{zphuang@cs.hku.hk}

% \author{Yudian Zheng}
% \affiliation{
%  \institution{The University of Hong Kong}
%  }
% \email{ydzheng2@cs.hku.hk}

%\pagestyle{plain}

\maketitle
%!TEX root = main.tex

\begin{abstract}

% Reynold 27/5
{\it Crowdsourcing} employs human workers to solve 
computer-hard problems, such as data cleaning, entity resolution, and sentiment analysis.
When crowdsourcing tabular data, e.g., the attribute values of an entity set,
a worker's answers on the different attributes 
(e.g., the {\it nationality} and {\it age} of a celebrity star) are often treated independently. This assumption is not always true and can lead to suboptimal crowdsourcing performance.  In this paper, we present the {\it T-Crowd} system, which takes into consideration the intricate relationships among tasks, in order to converge faster to their true values.
Particularly, T-Crowd integrates each worker's 
answers on different attributes to effectively learn his/her trustworthiness and the true data values.  
The attribute relationship information is also used to guide task allocation to workers. 
Finally, T-Crowd seamlessly supports categorical and continuous attributes, which are the two main datatypes found in typical databases.  
Our extensive experiments on real and synthetic datasets show that T-Crowd outperforms state-of-the-art methods in terms of truth inference and reducing the cost of crowdsourcing.   

% Before Reynold's edit on 27/5
% {\it Crowdsourcing} is the practice of
% obtaining information 
% %or input into a task 
% by enlisting the services of a large number of people, called {\it the crowd}. Existing work on crowdsourcing typically assumes independence between the data items acquired from the crowd, and focuses on a single datatype, either {\it categorical} or {\it continuous}. In this paper, we study the use of crowdsourcing to collect {\it structured data}, where the answers to tasks that refer to the same entity are correlated with respect to their quality.
% Existing work does not fully consider this correlation, %among micro-tasks, thus 
% failing to perform well in both truth inference and task assignment. We propose a unified framework for crowdsourcing multi-type structured data, which has two distinct features compared to previous work. First, we unify worker quality on categorical data and continuous data in a probabilistic model, which helps us to infer the truth from the answers more accurately. Second, we propose 
% %a potask assignment policy that
% a policy for assigning tasks to the crowd
% %{\em inherent}  and {\em correlation-aware} task assignment policy 
% that considers the estimated quality of workers and the correlation among tasks for data items that refer to the same entity. 
% %To evaluate our proposed model, 
% We perform extensive experiments on three real datasets, which show that our proposal significantly outperforms the state-of-the-art methods.

\end{abstract}
%!TEX root = main.tex

\section{Introduction}\label{sec:introduction}

Crowdsourcing is an effective way to address computer-hard problems 
(e.g., entity resolution \cite{Crowder, waldo, falcon} and sentiment analysis \cite{qasca, cdas}) 
by utilizing numerous ordinary humans (called {\em workers} or {\em the crowd}). The general workflow of crowdsourcing is as follows: at first a {\em  requester} 
 proposes a problem, then the problem is transformed into many 
tasks (i.e., questions), and finally the workers complete the tasks assigned to them and they are given a monetary reward. 
%A HIT includes a number of specific questions, which we call {\em micro-tasks} on one or more subjects.
%Based on the non-experts and paid tasks which is two typical characteristics of crowdsourcing, 

%A common and important usage of 
%The main goal of crowdsourcing is to collect data from the crowd. A 
Many applications\cite{crowddb, deco,crowdfill} crowdsource tabular data, 
%the data collected by crowdsourcing 
%are 
i.e., a collection of discrete and related items
% that may be accessed individually or in combination or managed as a whole entity. 
which are structured in a tabular form
and
%, present in a tabular form and commonly 
comply to a 
%traditional \emph{relational database} 
schema. Each column represents a particular attribute or variable. Each row corresponds to an entity and includes a value for each of the variables.
% given member of the dataset in question. It lists values for each of the variables. 
Table \ref{table:example_celebrity} illustrates an image recognition example; given the picture of a celebrity, the goal is to collect the name, nationality, age, and height of the person from the crowd. The values shown in Table \ref{table:example_celebrity} are the unknown (ground) truth data to be collected from the workers.
Each cell of this table can be considered as a task, i.e., a worker may be asked to provide a value for the name of a celebrity given his/her picture \cite{crowdfill}. 
%\nikos{cite some crowd-db papers here such as crowdfill or crowddb if they are consistent to this statement.}\caihua{Done}
%\nikos{it appears that there is an inconsistency here. in all the paper, you use `task' to mean one cell of the table and these tasks can be grouped to form HITs. However, in the beginning of this section, you use `task' to mean the whole crowdsourcing problem. If the common notation is to use task as the whole problem, i.e. the whole table, here you should define and use another term, for example `microtask' and you should clearly say that these microtasks are grouped into HITs and given to the workers.} \caihua{solved, see the first paragraph}
%Such tasks are common in  
%given a picture of celebrity. It's useful in image recognition. 
%Several characteristics define a dataset's structure and properties. 
%The structure of a crowdsourced dataset includes the number and types of the attributes or variables. 

\begin{table}[] %\vspace{-.5em}
\makegapedcells
%\caption{Ground Truth of Celebrity Info Given a Picture} \small
\caption{Ground Truth about Celebrities} \small%\scriptsize
\label{table:example_celebrity}
\setcellgapes{1pt}
\newsavebox{\mybox}
\newcolumntype{X}[1]{>{\begin{lrbox}{\mybox}}c<{\end{lrbox}\makecell[#1]{\usebox\mybox}}}
\begin{tabular}{X{cc}@{~}|X{cc}@{~}|@{~}X{cc}@{~}|@{~}X{cc}@{~}|@{~}X{cc}@{~}|@{~}X{cc}@{~}|} \cline{2-6}
 & Picture  &  Name  & Nationality & Age & Height  \tabularnewline \cline{2-6}
1 &{\includegraphics[width=0.07\textwidth ,height=0.05\textheight,bb=0 0 250 250]{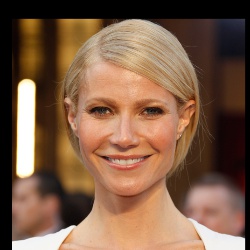}} & Gwyneth Paltrow         & United States &     40      &     5$'$9   \\ \cline{2-6}
2 &{\includegraphics[width=0.07\textwidth ,height=0.05\textheight,bb=0 0 250 250]{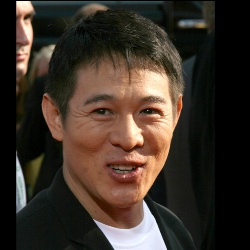}} & Jet Li            & China &     45      &     5$'$6    \\ \cline{2-6}
3 &{\includegraphics[width=0.07\textwidth ,height=0.05\textheight,bb=0 0 250 250]{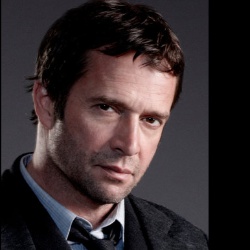}}&  James Purefoy & Great Britain &     48      &     6$'$1     \\ \cline{2-6}
\end{tabular} 
\end{table} %\vspace{-.5em}\vspace{-.5em} 

\begin{table}[] 
%\vspace{-1em}
\makegapedcells
%\caption{Answers of Celebrity Info Given a Picture} \small
\caption{Answers to Tasks about Celebrities} \small%\scriptsize
\label{table:example_celebrity_answers}
 \setcellgapes{1pt}
\newcolumntype{X}[1]{>{\begin{lrbox}{\mybox}}c<{\end{lrbox}\makecell[#1]{\usebox\mybox}}}
\begin{tabular}{|@{~}X{cc}@{~}|X{cc}@{~}|@{~}X{cc}@{~}|@{~}X{cc}@{~}|@{~}X{cc}@{~}|@{~}X{cc}@{~}|} \hline
Worker & Pic Id &  Name  & Nationality & Age & Height  \tabularnewline \hline

\multirow{2}{*}{$u_1$}    & 1 &  Gwyneth Paltrow  & United States & 39  & 5$'$9     \\ \cline{2-6} 
                 & 2 & Jet Li & China & 47  & 5$'$6     \\ \hline

\multirow{2}{*}{$u_2$} & 1 &  Gwyneth Paltrow  & Canada & 45  & 5$'$11     \\ \cline{2-6} 
 & 3 & James Purefoy & Great Britain & 51  & 6$'$     \\ \hline

\multirow{2}{*}{$u_3$} &2 &  Jet Li            & China &     45      &     5$'$6    \\\cline{2-6} 
&3 &  Ciarán Hinds & United States &     35      &     5$'$11     \\  \hline

%\multirow{2}{*}{w4} &2 &  Jackie Chan          & China &     45      &     5' 8  \\ \cline{2-6} 
%&3 &  Ciarán Hinds & United States &     35      &     5' 11     \\ \hline

\end{tabular}  %\vspace{-.5em} \vspace{-.5em} 
%\vspace{-2em}
\end{table}

Crowdsourcing involves two interrelated processes: \emph{truth inference} and \emph{task assignment}.
% become popular research topics nowadays. 
Truth inference refers to addressing noise and errors from the crowd in order to eventually infer the correct answer (or truth) for each task based on the answers to it by all workers \cite{em, yin2008truth}. Existing works \cite{ZenCrowd, GLAD, GTM, minimax} often model that a worker’s quality is consistent in all the tasks.
%Truth inference relies on developing a model that captures (i) the quality of each worker and (ii) the difficulty of each task. Given this model, the collected answers for the task are aggregated and the true value of the task is estimated\cite{GLAD,CRH,ZenCrowd,GTM}.
Task assignment refers to selecting an appropriate set of tasks to assign to each incoming worker. 
Selection of a task to assign 
can be based on 
%is typically based on two factors (i) 
how confident we are already about the true value of the task (i.e., whether we need more answers) \cite{askit} and/or 
%(ii) 
what is the estimated quality (i.e., reliability) of the worker on the specific task (i.e., the expected {\em information gain} after the corresponding assignment) \cite{qasca, CrowdDQS}. 
Truth inference can be used as a module in task assignment, 
to estimate the confidence of estimated true values  \cite{askit,cdas}.
% and the qualities of workers \cite{askit,cdas}. \nikos{cite some papers that do this}.\caihua{Solved.}
%\caihua{Similarly, not all methods use confident of the task and quality of the worker.}
%\nikos{but in related work and experiments you say that CDAS and Askit do not use the worker quality in task assignment. so, is there any related work that uses truth inference to estimate worker quality in task assignment????} 

Most crowdsourcing systems assume that the set of crowdsourced tasks are {\em homogeneous} and {\em independent} to each other. For example, consider an image tagging problem; all tasks request the same type of input from the workers (i.e., a set of keywords) and there are no obvious dependencies among different images.
%between the qualities of a worker's answers to different images.
%a subset of images that does not exist between the images of another subset. 
In this paper, we focus on crowdsourcing tabular data, which are directly related to database applications. 
We identify the properties of these data that 
make the application of previous work challenging and create opportunities 
for more effective
 crowdsourcing. 

First, the datatypes and domains of different attributes may vary. For example, in Table  \ref{table:example_celebrity}, the task ``What is the nationality of the person in Picture 1?'' has a different datatype compared to the task   ``What is the height of the person in Picture 2?'' (i.e., categorical vs. continuous). Even attributes of the same datatype may have different domains. As a result, approaches for integrating the answers of a worker in homogeneous tasks in order to estimate the worker's quality 
%(i.e., reliability)
% 
%\footnote{These include the popular EM algorithm \cite{emAlgorithm} for categorical data and data integration models applied for continuous attributes (GTM \cite{GTM} and CATD \cite{CATD})} 
are not directly applicable.
These include the popular EM algorithm \cite{emAlgorithm} for categorical data and data integration models applied for continuous attributes (GTM \cite{GTM} and CATD \cite{CATD}), to be discussed in Section \ref{sec:relatedwork}. As we will show, applying a different approach for each column does not transfer the knowledge from one datatype to the other, i.e., the estimation of worker quality can be inaccurate due to data sparsity. 

Second, in tabular data, there are potential dependencies between rows and columns.
%, e.g., with respect to the difficulty of tasks. 
The difficulty of a task might depend on the corresponding entity and attribute. As a result, the quality of a worker on a particular task may depend on his/her quality on other tasks in the same row or column. 
Take Table \ref{table:example_celebrity_answers} as an example, which contains the answers of three workers on tasks from Table  \ref{table:example_celebrity}. Observe that worker $u_3$ inputs a wrong name for the third picture, which means that he does not recognize this celebrity. If we allow him to provide values for the nationality, age and height of the person in that  picture, his answers would be unreliable, despite the high quality of his input for the second picture.
This means that when computing the quality of a worker for truth inference or task assignment, we should not consider the columns independently, but we should take such possible dependencies under consideration.

%\stitle{Contributions}
%To address these challenges, 
In this paper, we present the {\it T-Crowd} system, 
the first crowdsourcing solution 
that considers all the aforementioned 
%that provides a comprehensive framework for considering the listed 3 
properties of tabular data in both truth inference and task assignment. 
%(i.e., 
%In previous work, some of these properties are considered in either truth inference or task assignment, but not in both. 
%Specifically, 
T-Crowd 
%includes 
%a novel approach 
%that models the quality of workers on heterogeneous data in a uniform manner.
%We propose a novel method to uniform workers' quality on heterogeneous data explicitly. It 
%Our approach 
processes the submitted answers by each worker
%answers from a worker 
to infer a {\em single quality} for him or her, based on the assumption that the quality of each worker is consistent throughout the entire table.
T-Crowd seamlessly integrates the worker's answers to tasks of different datatypes and domains,  addressing consistency and data sparsity issues 
that would arise from the alternative approach of 
using different models for different columns.
For example, the overall quality of worker $u_1$ can be regarded as better than that of worker $u_2$ considering their answers to both categorical and continuous values in Table \ref{table:example_celebrity_answers}.  
Unified worker quality greatly improves the accuracy of truth inference and the performance of task assignment, reducing the total number of tasks to be assigned to workers until all true values are estimated with high confidence.

T-Crowd includes a model that 
captures the importance of tasks (i.e., how confident we are about their value estimates)
in the different columns and rows, based on the collected data so far.
%, since we are collecting record data having multiple datatypes.  
%even with different types of data. 
We also consider the quality of the worker who is answering and define an {\em inherent information gain} which is a uniform measure 
for ranking tasks with respect to a given worker.
%of obtaining additional answers for a certain task. 
Then we choose to assign to the worker the tasks with the highest anticipated benefit. 
%Although they are not in the same row or the same column which is not interface friendly to workers, it maximizes the utility of budget.
In contrast, previous work on 
crowdsourcing tabular data
%solving complex tasks involving multiple attributes in a table
%\cite{crowddb,deco,crowdfill}
%\noindent \textbf{Challenges in Task Assignment.}
%\stitle{Challenges in Task Assignment}
%Recently, crowdsourcing has been used to solve complex tasks involving multiple attributes in a table
%\cite{crowddb,deco,crowdfill}.
% a tabular setting. 
%CrowdDB\cite{crowddb} and Deco\cite{deco} assign a microtask-based HIT to obtain missing information in a relational database. CrowdFill\cite{crowdfill} randomly shows a partially-filled part of a table to workers and these workers are asked to change this part using three operations: fill in some empty cells, upvote or downvote rows entered by other workers. However, all these approaches do not consider that workers may have different quality in different HITs; for example, spammers or vicious annotators can seriously compromise the quality of a crowdsourcing system.
%will cause the crash of the whole system without quality control. 
performs task assignment based on only 
how many more answers are needed for each task, 
disregarding worker quality.
%before the true answer can be estimated with high confidence.
%However, this disregards the potentially different  benefit of obtaining additional answers for different tasks. 
%Hence, collecting additional answers to simple tasks for which 
%good results are obtained already would cost us budget that can be allocated to hard tasks that need more answers. 
To further improve performance, we utilize the potential correlations between tasks.
%Take Table \ref{table:example_celebrity_answers} as an example. Worker $u_3$ gives a wrong name to the third picture, which means that he does not recognize this celebrity. If we allow him to provide values for the nationality, age and height of the third picture, then his answers would be unreliable, despite the high quality of his input for the second picture.
%To address this issue, 
We define a {\em structure-aware information gain} which extends the inherent information gain to also 
consider as a parameter 
%which highlights 
the previous answers given by the worker on tasks that appear in the same row, when selecting new tasks to assign to him or her. 

To summarize, our main contributions are as follows:
\begin{itemize}
%\noindent $\bullet$ 
\item
We unify worker quality for all tasks in crowdsourced tabular data, improving
the accuracy of truth inference and the performance of task assignment, compared to models that treat each attribute independently. 
%addressing consistency and data sparsity issues. 
%Unified worker quality improves the accuracy of truth inference and the performance of task assignment.
% explicitly. We also design two factors to model the difficulty of structured data. 
%\noindent $\bullet$ 
\item Given an incoming worker, we find a suitable set of tasks for him/her based on the benefit of obtaining additional answers in tasks, the worker's inherent quality, and the correlation of answer quality between tasks in the same row.
%\noindent $\bullet$ 
\item
%We evaluate T-Crowd on three real-world datasets; the results demonstrate its superiority over existing alternatives. Compared to previous work, T-Crowd has superior truth inference accuracy and converges to the true answers of the tasks after a smaller number of task assignments.
We evaluate T-Crowd on real and synthetic datasets; the results demonstrate its superiority over existing alternatives. Compared to previous work, T-Crowd has better truth inference accuracy and converges to the true values of the tasks using only about half of the answers by the workers. 

%in collecting heterogeneous structured data on effectiveness both in truth inference and task assignment. It improves existing methods due to the uniform worker quality and correlation relationship in the table.
\end{itemize}

%In the rest of the paper,
The rest of the paper is organized as follows.  
%Related work is discussed in 
Section \ref{sec:relatedwork} discusses related work.
Section \ref{sec:problem} defines the problem and gives an overview of our system. In Section \ref{sec:TruthInference}, we present our methodology for truth inference. Our task assignment policy is presented in Section \ref{sec:Taskassignment}. Section \ref{sec:experiments} includes our experimental evaluation. Finally, we conclude in Section \ref{sec:conclusions}.

%!TEX root = main.tex

\section{related work}\label{sec:relatedwork}
Related work falls into two categories: {\em truth inference} methods used to infer the truth
%of crowdsourced tasks based on the collected answers by the workers (which can be noisy) 
and {\em task assignment} strategies for a incoming worker. 
 
%To obtain data of high quality at a low cost and latency, crowdsourcing systems use 
%two main strategies to control: 
%{\em truth inference} methods and {\em task assignment} strategies. 

%\noindent \textbf{Truth Inference.} 
\stitle{Truth Inference}
%Truth inference combines the answers collected from the workers (which can be noisy) to infer the true answer, by 
%workers' answers to detect the truth by 
%estimating the reliability of each worker. 
The most basic truth inference methods are majority voting for multiple-choice tasks (i.e., categorical data) and taking the median for numerical tasks (i.e., continuous data). 
These approaches regard all workers as equal, disregarding any differences in their trustworthiness. 
Methods such as D$\&$S \cite{emAlgorithm,panos_qualitymanagement} use a confusion matrix to model a worker's quality of a worker, and then use an Expectation-Maximization (EM) algorithm to infer the truth.  More advanced approaches like TruthFinder~\cite{TruthFinder}, Accusim~\cite{Accusim}, and GLAD~\cite{GLAD} improve accuracy using different worker answering models or by considering more parameters, such as the difficulty of the task. These methods focus on answering tasks on categorical data. Other methods, such as GTM~\cite{GTM}, are designed for continuous crowdsourced data. CRH~\cite{CRH,CRH2} and CATD~\cite{CATD} are two existing truth inference approaches for both categorical and continuous data.
%truth discovery approaches suitable for both categorical and continuous data. Truth discovery is similar to inference in  crowdsourcing; if the input sources (e.g., websites) can be regarded as workers then CRH~\cite{CRH} and CATD~\cite{CATD} can be applied directly in crowdsourcing. 
CRH~\cite{CRH} incorporates different distance functions between the answers and the estimated truth to recognize the characteristics of various data types. Specifically, CRH proposes an objective function and minimizes it by updating the estimated true values and source reliability (i.e., worker quality) in turns. CATD~\cite{CATD} considers both source reliability and the confidence interval of the estimation. 
Additional information of tasks or workers has also been considered in truth inference, such as the latent topics of the tasks \cite{faitcrowd} and the learn bias of workers \cite{debias}.
%In other 
%More inference situations have been considered in~\cite{faitcrowd, Debias} using additional information of tasks or workers. For example, \cite{faitcrowd} uses the content of tasks to decide domains of tasks and \cite{debias} needs some training data and learn bias of workers.
%{\color{red} [Rey: not clear what are ``inference situations'']}\caihua{It's about using additional data, like profile of workers, to infer something.}
%\nikos{say EXACTLY what is used by each of \cite{faitcrowd} and \cite{debias}}

The aforementioned works do not take into account the nature of tabular data that we address in this paper.
In Section \ref{sec:TruthInference}, we present an iterative Expectation-Maximization (EM) truth inference algorithm, which improves
% which is related to EM solutions, 
%which update its parameters until an objective function converges. 
%a novel method to unify the workers' quality on heterogeneous tasks (i.e., continuous and categorical attributes) and
%improve 
the accuracy of truth inference from the answers compared to previous work. The 
novelty of our work
%main difference between our method and CRH 
is that we use a probabilistic model for the answers of workers for different data types and that we unify workers' quality on categorical data and continuous data explicitly, while methods like CRH design different distance functions for the different data types.
As CRH is primarily designed to discover truth from web data, it may not adapt well to the problem of inferring truths from workers' answers, which exhibit a long-tail distribution. Our approach, which is customized for crowdsourcing, performs better than CRH experimentally. 
%
%
%Our approach is expected to be more accurate in our problem setting compared to CRH, because 
%CRH is designed Truth Discovery from web data.
%Websites are more trustworthy while workers' quality is more random. In addition, for the different tasks the number of answers given by the workers follow a long-tail distribution (i.e., data are more sparse).

%\noindent \textbf{Task Assignment.} 
\stitle{Task Assignment}
Online task assignment 
selects which tasks to assign to each incoming worker,
%matches tasks and workers
in order to achieve the maximum possible quality for the collected data.
% which influences the quality and cost of crowdsourcing system. 
%There are two scenarios: one is worker-based \cite{OptKG, bandit} which is selecting worker and task together and other is task-based where we only can select tasks for certain next worker. In this situation, we don't know which worker will come next time and the number of questions a worker will answer. Because on commercial platforms, like AMT\cite{AMT} and CrowdFlower\cite{crowdFlower}, we can't control workers' behavior, we focus on the task-based situation.
In simple crowdsourcing systems, like CDAS~\cite{cdas}, the candidate tasks are randomly assigned to workers. 
AskIt~\cite{askit} is yet another crowdsourcing platform, which assigns the tasks that have the highest uncertainty, again disregarding the quality (or expertise) of the incoming worker for these tasks. 
%Crowdsourced 
%Query processing systems for crowdsourced data, such as 
CrowdDB~\cite{crowddb}, Deco~\cite{deco}, and Qurk~\cite{qurk} are 
extensions of relational database systems that incorporate the crowd's knowledge into query processing. 
They use answers from the crowd to make up the missing values of query operators. 
They are similar to our approach in that they collect tabular data; however, they do not focus on the assignment strategy and simply assign random tasks to workers. 
CrowdFill~\cite{crowdfill} is the most similar and recent system for tabular data. In CrowdFill, the workers are asked to select and answer tasks from a subset of the table given to them and they can also vote for the answers to these tasks by other workers. Due to the different way of acquiring data from workers, CrowdFill is not directly comparable to our work. Besides, CrowdFill does not estimate worker quality, and does not use properties of tabular data (e.g., attribute dependencies) to assign tasks to workers. 
%It is interesting to investigate how our task assignment solutions, designed for tabular data, can improve the task assignment process of CrowdFill.
%should be directed to which workers, based on only 
%, in order to reduce the uncertainty about the collected data in multiple-choice tasks.
%QASCA~\cite{qasca} selects to assign the tasks that have the highest expected improvement in quality, using Accuracy and F-score metrics. 
%\nikos{unclear if worker quality is also considered. unclear why you do not compare to this method.}
Some methods~\cite{icrowd, docs,www_hierarchy_taskassign} consider the case where the tasks are relevant to different domains and workers are given the tasks that match their domain expertise. In~\cite{OptKG, bandit}, task assignment is modeled by a Markov Decision process and solved as a multi-armed bandit problem, but the application of the model is limited to only single-choice or multiple-choice tasks. Other forms of online task assignment, which need explicit workers' collaboration, have been studied in~\cite{collaborative1, collaborative2}.

Different from the above works, our method focuses on crowdsourcing datasets, which are structured and heterogeneous, presenting  challenges and opportunities as discussed in the Introduction. 
%Therefore, data collection cannot be simply transformed into a number of decision-making (`T' or `F'), multiple-choice or numeric tasks because we want to obtain global optimum. \nikos{unclear. what do you mean by `global optimum'?} \caihua{`global optimum' means that we get estimated truth of each cell in the whole table under certain budget. `Global' means whole table. If we consider each column separately, how to divide budget into each column and how to compare uncertainty of multiple-choice or numeric task are hard.}
We consider the currently collected data, the difficulty of tasks, and the correlations of answer quality for tasks that refer to the same entity, to 
estimate the quality of workers and to conduct task assignment targeting the maximization of information gain of tasks.
%etc.} \caihua{Yes!} to estimate workers' personalized and correlated quality in structured data \nikos{unclear}. Intuitively, our method selects a suitable task to the worker for which the expectation of reduced uncertainty \nikos{what do you mean by `uncertainty'? do you mean `expected error'?} \caihua{I explain it in the first paragraph in section task assginment.} in the whole data is maximum based on his predicted answer for this task.

%!TEX root = main.tex

\begin{table}[!ht]
\centering
\caption{Table of Notations} \small%\scriptsize
\label{table:notation}
\begin{tabular}{|@{~}c@{~}|@{~}c@{~}|}
\hline
\textbf{Notation} & \textbf{Description} \\ \hline \hline
$c_{ij}$        	&	cell (task) in the $i$-th row and $j$-th column     \\ \hline
$a_{ij}^u$         &	answer given by worker $u$ for cell $c_{ij}$       \\ \hline
$\mathcal{A}$ &  the set of all answers, i.e., $\mathcal{A}=\{a_{ij}^u\}$  \\ \hline
$T_{ij}$         	&  	distribution of estimated truth for cell $c_{ij}$           \\ \hline
$T_{ij}^{*} ~(\widehat{T}_{ij})$         	&  ground truth (estimated truth) for cell $c_{ij}$           \\ \hline
$e_{ij}^{u}$ & error of $a_{ij}^u$ with respect to $\widehat{T}_{ij}$ \\ \hline
$ q_u$ 		&      quality of worker $u$    \\ \hline
%$ q_u (\phi_u) $ 		&      quality of worker $u$    \\ \hline
$ \alpha_i ~(\beta_j)$ & difficulty of row $i$ (column $j$)  \\ \hline
\end{tabular}
\vspace{-0.7em}
\end{table}

\section{Problem Definition}\label{sec:problem}

In this section, we formulate the problem and give an overview of T-Crowd. 
%For similarity, Table \ref{table:example_celebrity} is used as an example for illustration.
%\subsection{Problem Formulation}
As discussed in Section~\ref{sec:introduction}, our goal is to perform crowdsourcing on a two-dimensional table $C$, defined as follows.

\begin{definition}[Tabular Data Model]
We target the crowdsourcing of a two-dimensional table $C = \{c_{ij}\}$, where $i \in \{1,...,N\}$ and $ j \in \{1,...,M\}$. $C$ has an {\it entity} attribute which is the key attribute of the table. Each column is a categorical or a continuous attribute. Each cell $c_{ij}$ represents the value of the $i$-th entity in the $j$-th attribute, whose true value (i.e., \emph{truth}, or \emph{ground truth}) is denoted as $T_{ij}^{*}$.  
\end{definition}

Table~\ref{table:example_celebrity} shows an example of tabular data about celebrities that we want to crowdsource. 
\emph{Age} and \emph{Height} are continuous attributes, while \emph{Name} and \emph{Nationality} are categorical attributes. The entity attribute is \emph{Picture}. 
To obtain the truth for the remaining attributes, we ask the crowd to provide answers.

\begin{definition}[Task, Worker, Answer] 
A task is \\related to a cell $c_{ij}$ and the crowd (or workers) is asked to answer the task, by providing values for the cell. Let $U$ be a set of workers.
A worker $u\in U$ will submit an answer $a_{ij}^{u}$, if cell $c_{ij}$ is assigned to $u$.
\end{definition}

For example, to get the age of the second entity, a task provides the picture of the second entity and asks workers to input the age. 
Since workers may have different levels of quality (e.g., some workers are experts, while some are spammers), each task $c_{ij}$ is often assigned to multiple workers and all acquired answers for $c_{ij}$ are aggregated to infer the true value of $c_{ij}$. 
Next, we define the two problems that we are addressing in this paper.

\begin{definition} [Truth Inference]
Given the set of \\answers $\{a_{ij}^u\}$, by workers $u \in U$
to cells $c_{ij}$, $i \in \{1,...,N\}$, $j \in \{1,...,M\}$, 
the problem of truth inference is to compute an accurate estimate $\widehat{T}_{ij}$ for each cell $c_{ij}$'s true value $T_{ij}^{*}$.
\end{definition}

% \begin{definition} [Task Assignment Problem]
% When a worker $u$ request tasks for $C$, the task assignment problem determines which $K$ tasks to assign to $u$, based on the  previous answers by $u$ or other workers on tasks of $C$.%
% \footnote{Existing crowdsourcing platforms, such as Amazon Mechanical Turk (AMT)~\cite{AMT}, often support the functionality of dynamic assigning a set of tasks to a coming worker (e.g., the `external-HIT'~\cite{externalHIT} way in AMT).} 
% % When a worker $u$ requests a HIT, the problem of task assignment is select the best possible cells to include in $u$'s HIT, considering the anticipated reduction in the true value uncertainty of these cells after $u$'s answer.
% %to set up a optimal HIT $H_{u}$ for $u$ based on the current confidence of the estimated truth $T_{ij}$ and his quality $q_u$ from answer history adaptively.
% \end{definition}

\begin{definition} [Task Assignment]
When a worker $u$ requests for a task for $C$,  decide the 
%$K$ 
task to be assigned to $u$.
% based on $u$'s previous answers by $u$ or other workers on tasks of $C$.%
%\footnote{Existing crowdsourcing platforms, such as Amazon Mechanical Turk (AMT)~\cite{AMT}, often support the functionality of dynamic assigning a set of tasks to a coming worker (e.g., the `external-HIT'~\cite{externalHIT} feature in AMT).} 
% When a worker $u$ requests a HIT, the problem of task assignment is select the best possible cells to include in $u$'s HIT, considering the anticipated reduction in the true value uncertainty of these cells after $u$'s answer.
%to set up a optimal HIT $H_{u}$ for $u$ based on the current confidence of the estimated truth $T_{ij}$ and his quality $q_u$ from answer history adaptively.
\end{definition}

%\nikos{the definition above is problematic. $T_{ij}$ seems to be out of place. $T_{ij}$ refers to a specific cell. Normally you should say that the HIT $H_u$ should include cell $c_{ij}$ if the confidence of $T_{ij}$ is low and also the quality of the worker is estimated to be high. Also, ``from answer history adaptively'' is not accurate. In definitions, you should well defined wordings.}\caihua{

%(Task Assignment). When a worker $u$ requests a HIT, the problem of task assignment is to select $k$ cells $c_{ij}$ to set up a HIT $H_{u} = \{c_{ij}\}_{k}$. 

%As for how to select these cells, in this paper we consider some measures, such as uncertainty (information gain) of each cells, worker quality and correlation.

%}

As we will discuss, a worker's previous answers, as well as other workers' answers, are both instrumental in task assignment. It is also worth to note that existing crowdsourcing platforms, such as the Amazon Mechanical Turk (AMT)~\cite{AMT}, support the functionality of dynamically assigning tasks to an incoming worker (e.g., the `external-HIT' feature in AMT~\cite{externalHIT}). Table \ref{table:notation} summarizes the notations used in this paper.

\begin{figure}[!t] %\vspace{-2em}
 \centering
 \includegraphics[width=1\columnwidth]{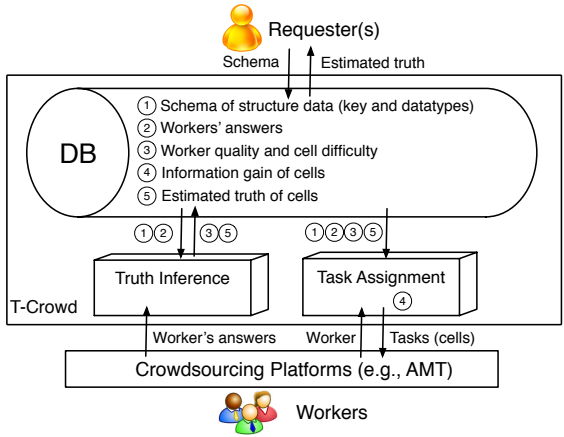}
 \caption{System Architecture}\label{fig:Workflow} \vspace{-1em}
\end{figure}

%\subsection{System Architecture}
\stitle{System architecture} 
Figure \ref{fig:Workflow} gives an overview of T-Crowd, 
our proposed system for crowdsourcing tabular data.
A {\it requester} (e.g., a lifestyle journal) first defines the structure (i.e., schema) of the collected data, such as the datatypes of the columns, and the key attribute. 
Then the requester publishes tasks to a crowdsourcing platform, e.g., AMT~\cite{AMT}. For an incoming worker $u$, 
our Task Assignment module 
%${\tt task~assignment}$ 
determines one or more cells and assigns the corresponding task(s) to 
%generates a set $D$ of $K$ cells and assigns $D$ to 
$u$. This is based on the anticipated {\it information gain} of the different cells by $u$'s answers.
Intuitively, the information gain is an estimate of how much more accurate the cells' values become upon collection of $u$'s inputs. 
When the worker submits an answer $a_{ij}^{u}$ for a cell $c_{ij}$ to the system, %${\tt truth~inference}$ 
the Truth Inference module infers the estimated truth $\widehat{T}_{ij}$. To facilitate task assignment and truth inference, we also estimate the quality of worker $q_u$ and the difficulty of cells $\alpha_i$ and $\beta_j$. 
It assigns task(s) to workers and collect answers from workers before the budget is exhausted. 
%Next, we present in detail  the truth inference module of T-Crowd.

%The next section presents the details.

%\reminder{LGL: add the symbols, e.g.,  $q_u$,$\alpha_i$ and $\beta_j$ in the figure?}

% In our crowdsourcing problem, we also define the number of questions  $k$ (i.e., cells) that are asked together in each HIT and a total 
% %The requester is also supposed to indicate the number of cells in each HIT $k$ and total 
% budget $B$ (if each cell costs the same amount of money $b$, then the total number of HITs that will be given to workers is $\frac{B}{k*b}$).
% % \nikos{give a precise definition of budget. the optimization problem is unclear here. is this the total number of HITs?} \caihua{each cell costs the same money $b$. The number of HITs is equal to $\frac{B}{k*b}$.}
% %to ensure the total number of HITs and cells included in a HIT. 
% %Note that a worker may request a HIT multiple times, therefore we have the history of assigned HITs and his answers. 
% The system sets up a HIT for a worker based on 
% the current uncertainty about the true values of cells and the anticipated quality of the worker 
% according to his answers at previous HITs.
%demand for more answers for a cell 
%confidence of evaluated truth given the current data and the worker's quality according to the history of answers.
%\nikos{explain `adaptively'. Is it the history of the same worker or all workers?} \caihua{Maybe `adaptively' is not suitable. Delete it?}

%!TEX root = main.tex

\section{Truth Inference}\label{sec:TruthInference}

In this section, we explain how T-Crowd performs truth inference on tabular data.
%, which addresses the consistency issue in
%the quality of workers across different columns and the 
%varying difficulty in answering different cells. 
The quality of truth inference for a data cell $c_{ij}$ depends on the
quality of workers who answer $c_{ij}$, 
and the difficulty of $c_{ij}$. We first discuss how to model 
worker quality and cell difficulty (Sections~\ref{subsec:workermodel}
and \ref{sec:difficulty of a cell}). Then, we show how to infer
the true values of cells based on these two factors (Section~\ref{subsec:inference}).
%To integrate the qualities of a worker in different columns, our idea is that the average error of a worker in two different columns are positively correlated (which we will show in Section~\ref{sec:experiments}. That is, if a worker answers wrongly in one attribute of an entity, he/she would be highly possible to answer incorrectly for the other attributes of that entity. 

% . As observed from our empirical
% investigation in Section \ref{sec:experiments}, the average error
% of a worker in two different columns (no matter whether the data
% types of these columns are the same) are positively correlated.
% This means that if a worker answers wrongly in one attribute of an
% object, we can infer that she would probably provide an incorrect
% answer for the other attributes of that object. 
%Since most of existing works employ one parameter to evaluate the quality of each worker based for each attribute (one column in the table),
%Motivated by this observation,  we use a single parameter $q_u$ to model the quality of a worker $u$,  which can be used uniformly on all attributes.
%to serves as a means for uniformly estimating 
%each worker's qualities based on all attribute of the objects (all
%columns in the table). 
%

%\subsection{Worker Model}\label{subsec:workermodel}
\subsection{Quality of a Worker} \label{subsec:workermodel}

The challenge in modeling worker quality is that attributes may have
different datatypes and difficulties; the answer set of a categorical
task is finite and nominal, while that of a continuous task is an integer or a real number.
%\nikos{note that we can also have ordinal attributes that take numerical or other values from a finite, ordered set. That is a small set of ordered numbers (e.g., rank or rating is such a variable; age can also be considered such an attribute because the domain is finite - integers within a small range). In order to avoid questions, I suggest to rephrase this and avoid mentioning that continuous attributes have infinite values, otherwise the study seems incomplete.}\caihua{Yes, our method can deal with ordered numbers as continuous data. Age and notibiliity of celebrity is ordered but I regard them as ordered.}
Hence, it is not straightforward to model the quality of a worker using a single parameter. To address this problem, we propose a unified model for both categorical and continuous attributes. 

%\subsubsection{Data Types}

%Generally speaking, the datatypes of attributes can be divided in two types: \emph{categorical} and \emph{continuous}. The main difference is that answer set of a categorical task is finite and nominal, while that of a continuous task is an ordinal subset of $\mathbb{R}$. 
%Meanwhile, there exists a magnitude to estimate the closeness between workers' answers and the truth.
% \nikos{unclear. rephrase.} \caihua{Changed}

We model the truth of a categorical attribute $l^*$ as an element in a
finite unordered set of possible answers $L=\{l_1,l_2,...,l_{|L|}\}$. 
%\nikos{bad  notation. use $|L|$ for the number of possible answers. Also, $L$  and $V$ do not have different notation for different attributes. You  could use $L_a$ or $V_a$ for attribute $a$}\caihua{Changed}
% and each value has no magnitude.
%When a worker gives an answer, 
An answer from a worker is either correct or wrong
depending on whether it is the same as the ground truth.  
On the other hand, for a continuous attribute, 
the quality of the answer depends on how close it is to the ground truth. % and on the magnitude of this value.
%there is a magnitude to calculate the
%distance between the answer and the truth. 
For example, if the
height of Jet Li is 5$'$6, 
and a worker answers 5$'$7, which is close to the truth, the answer is
considered to be a good one.
%one answer is 1 billion and the
%other answer is 1.3 billion. We can say that the latter one is closer
%to the truth, compared with the former one. We can also infer that the
%quality of the latter worker should be estimated higher than the
%former one based on this observation. \nikos{the discussion is too
%  informal. There should be a precise definition, otherwise, skip it
%  and define it later precisely. Abstract discussions are not helpful
%  in this place.} 

%\nikos{the discussion above about categorical and continuous
%  attributes should be supported by citations to previous work, not
%  necessarily in crowdsourcing.}
%\caihua{These two data types are used in \cite{CRH}. It introduces in `introduction' but doesn't have a clear definition.}

As discussed, our goal is to use a single parameter $q_u$ to
represent the quality of a worker $u$. For the ease of presentation,
we first illustrate how the worker's quality for continuous datatypes
can be modeled, and then show how the model can be extended for categorical datatypes. 

% \begin{equation} \label{eq:definequalitycat}
% q_{u} =  P(a_{ij}^{u} = x ~|~ T_{ij} = x),~\text{where }~x\in L_j.
% \end{equation}
%\triangleq

%In other words, $u$ correctly answering a cell can be regarded as a Bernoulli distribution,

\noindent $\bullet$ For {\bf continuous} datatypes, we model the distribution of the answer given by worker $u$ as a normal distribution: $a_{ij}^{u} \sim \mathcal{N}(\widehat{T}_{ij}, \phi_{u}) $: 
%\nikos{symbol mismatch with section 2: either use hat or widehat, but not both} \caihua{Solved}
% using variance $\phi_{u}$ to estimate the quality of worker ${u}$. 
\begin{equation} \vspace{-.5em}
\label{eq:contap}
P(a_{ij}^{u} = x) = \frac{1}{\sqrt{2\pi\phi_{u}}}\, \exp\big({-\frac{(x - \widehat{T}_{ij})^2}{2 \phi_{u}}}\big), 
\end{equation}
where $\widehat{T}_{ij}$ is the expected value of $c_{ij}$ and $\phi_{u}$ is the variance of $u$. Intuitively, the higher
the quality of a worker is, the smaller the variance will be, as her
answer should have smaller difference from the truth. 
Inspired by this, we model $q_u\in [0,1]$ as the probability that the answer from
worker $u$ falls into a small range ($\epsilon$) around the truth $\widehat{T}_{ij}$:
%Since we know one worker quality is correlated in different columns, it's reasonable to build a bridge to estimate  $q_{u}$ and $\phi_{u}$  together. The variance $\phi_{u}$ of normal distribution means instability that a worker answers correctly. If a worker's answer of continuous attribute is just a little different from the ground truth, it means that his quality is stable and his answer of categorical attribute is correct probably because it's discrete. So we can use area of normal distribution curve in certain centre range to uniform $q_{u}$ and $\phi_{u}$. 
\begin{equation}
\label{eq:intial_trans}
q_{u} = P(~a_{ij}^u \in [\widehat{T}_{ij} - \epsilon, \widehat{T}_{ij} + \epsilon]~) = 
%\int_{-\epsilon}^{\epsilon}  \frac{1}{\sqrt{2\pi\phi_{u}}}\, e^{-\frac{x^2}{2 \phi_{u}}} \,dx =
 \text{erf}( {\epsilon}/{\sqrt{2\phi_{u}}}).
\end{equation}

\noindent Intuitively, $q_{u}$ is the area under the normal distribution curve, where
$\epsilon$ is a general parameter that controls the shape of the area and ``erf'' is the Gauss error function~\cite{gauss_error}. 
%Typical values of $\epsilon$ are between 0 to 3. \reminder{why 0 to 3??}

\noindent $\bullet$ For {\bf categorical} attributes, $q_{u} \in [0, 1]$ indicates the probability that the worker $u$ would correctly answer a task, i.e., 
\begin{equation}
\label{eq:cateap}
P(a_{ij}^{u} = z) = (q_u)^{\indfunc{\widehat{T}_{ij}=z}} \cdot \big(\frac{1-q_u}{|L|-1}\big)^{\indfunc{\widehat{T}_{ij}\neq z}}, 
\end{equation}
\noindent where $\indfunc{\cdot}$ is an {\em indicator function} which returns 1 if the argument is true; 0, otherwise. For example, $\indfunc{5=5}=1$ and $\indfunc{5=3}=0$. Intuitively, worker $u$ has probability $q_u$ to give the correct answer and we evenly distribute the probability $(1-q_u)$ to the remaining (false) answers. Note that $q_u$ can be expressed as in Eq.~\ref{eq:intial_trans}, meaning that we can use the same quality measure for categorical and continuous attributes.

%\zhipeng{Add discussion on $\epsilon$}
%Some empirical proof can be also found in Section \ref{sec:experiments}.

\subsection{Difficulty of a Cell} \label{sec:difficulty of a cell}
%In general, 
The answers from workers do not only depend on their
expertise, but they are also influenced by the difficulty of
tasks. 
% In our model, the difficulty of each cell $c_{ij}$ depends on the
% difficulty of attribute $j$ and the difficulty of object $i$. 
% \nikos{this is a sort-of independence assumption. You should
%   support the hypothesis.} \caihua{I know it's a hypothesis. I add two multipliers to represent difficulty like what is assumed in \cite{GLAD}. He also used multiplication.}
Hence, in our model, the quality of answer $a_{ij}^u$ depends on the quality of worker $u$, the difficulty $\beta_j$ of attribute (i.e., column) 
$j$, 
and the difficulty $\alpha_i$ of entity (i.e., row) $i$.

%We model the difficulty from the column and the row using the
%parameters $\alpha$ and $\beta$ separately. 
% Since a cell $c_{ij}$ includes the value of object $i$ in attribute
% $j$, 
% %is about one attribute $j$ of a object $i$, 
% the difficulty of a cell can be modeled as a function of the
% difficulties of the $i$-th row $\alpha_{i}$ and the $j$-th column
% $\beta_{j}$. 

To incorporate the difficulty of each cell $c_{ij}$ into the worker's quality,
% we model the finer granularity quality for each worker. That is, for a worker $u$, 
we define the variance of his/her answer to a cell $c_{ij}$ as $\phi^{u}_{ij} = \alpha_{i}\beta_{j}\phi_{u}$.
This means that the variance 
%that worker $w$ answers $c_{ij}$ 
is positively correlated to the difficulties $\alpha_i$ and $\beta_j$, and the inherent variance ($\phi_u$) of answers by worker $u$. 
Then, following Equation~\ref{eq:intial_trans}, we represent the quality of worker $u$ answering cell $c_{ij}$ as $q^{u}_{ij} = \text{erf}\left( \epsilon / \sqrt{2\alpha_{i}\beta_{j} \phi_{u}}\right)$. 
To model the worker's answers on categorical and continuous data, 
%Similarly, for worker model of continuous and categorical datatypes, 
Equations \ref{eq:contap} and \ref{eq:cateap} can be changed accordingly, i.e., by replacing $\phi_u$ with $\phi^u_{ij}$ and $q_u$ with $q^u_{ij}$.

Note that $\widehat{T}_{ij}$, $\alpha_i$, $\beta_j$ and $\phi_u$ are unknown and we discuss how to compute them later. The worker quality $q_u$ ($q_{ij}^u$) can be calculated directly if we know 
$\alpha_i$, $\beta_j$, and $\phi_u$.
%$\alpha_i$, $\beta_j$ and $\phi_u$ and we use it for simplicity.

% \begin{equation} \vspace{-.5em}
% \label{eq:contap}
% P(a_{ij}^{u} = x) = \frac{1}{\sqrt{2\pi\phi_{u}}}\, \exp\big({-\frac{(x - T_{ij})^2}{2 \phi_{u}}}\big), 
% \end{equation}

% \begin{equation}
% \label{eq:cateap}
% P(a_{ij}^{u} = x) = (q_u)^{\indfunc{T_{ij}=x}} \cdot \big(\frac{1-q_u}{|L|-1}\big)^{\indfunc{T_{ij}\neq x}}.
% \end{equation}

% Accordingly, formula \ref{eq:cateap} will be
% \begin{equation}
% P(a_{ij}^{u} = x) = 
% \begin{cases}
% q^{u}_{ij} &,~T_{ij} = x\\
% 1-q^{u}_{ij} &,~T_{ij} \not= x
% \end{cases}  \\
% \end{equation}
% and formula \ref{eq:contap} becomes
% \begin{equation}
% P(a_{ij}^{u} = x) \sim \mathcal{N}(T_{ij}, \alpha_{i}\beta_{j}\phi_{u})
% %&= \frac{1}{\sqrt{2\pi\phi^{u}_{ij}}}\, e^{-\frac{(x - T_{ij})^2}{2 \phi^{u}_{ij}}}  \\
% %&=  \frac{1}{\sqrt{2\pi\alpha_{i}\beta_{j} \phi_{u}}}\, e^{-\frac{(x - T_{ij})^2}{2 \alpha_{i}\beta_{j} \phi_{u}}} 
% \end{equation}

\subsection{Inference Process} \label{subsec:inference}
% observed variable $a_{ij}^{u}$ which is the answer from worker ${u}$ given cell $c_{ij}$ ,  hidden variable is the ground truth $T_{ij}$ and parameters are the difficulty of row ${i}$ $\alpha_{i}$, the difficulty of column ${j}$ $\beta_{j}$, and worker ${u}$'s quality $q_{u}$ and $\phi_{u}$. 

% \reminder{revise this section by:
% (1) clearly define each notation before using it (important), and try to simplify each formula; 
% (2) express your objective function (i.e., Eq~4); 
% (3) briefly illustrate the solution to the EM algorithm;
% (4) give an algorithm structure.}

The objective function of the truth inference problem is to maximize the likelihood of workers' answers, i.e., 
\begin{equation} \nonumber
\begin{aligned}
\arg\max  _{\alpha, \beta, \phi} P(\mathcal{A}|\alpha, \beta, \phi) = \arg\max  _{\alpha, \beta, \phi}  \sum\nolimits_{\mathcal{T}} P(\mathcal{A},\mathcal{T}|\alpha, \beta, \phi), 
\end{aligned}
\end{equation}
where $\mathcal{A}$ is the current set of answers by all workers on all cells and $\mathcal{T}$ is a set of all hidden true values, i.e., $\mathcal{T}=\{T_{ij}\}$. $T_{ij}$ denotes the estimated distribution of truth in cell $c_{ij}$.
%\nikos{$T_{ij}$ without a hat or star is not defined before}\caihua{Solved}
%\reminder{how to comute $P(\mathcal{A},\mathcal{T}|\alpha, \beta, \phi)$ is not defined here} 
To optimize this non-convex function, we use the Expectation-Maximization (EM) algorithm~\cite{em}, which takes an iterative approach.
%, while in each iteration, E-step and M-step are run, respectively. 
In each iteration of EM, the E-step computes the hidden variables in $\mathcal{T}$, and the M-step computes the parameters $\alpha_{i}$, $\beta_{j}$ and $\phi_{u}$ ($q_{u}$). Next, we provide details about the E-step and the M-step. 

% To start, we define the joint posterior distribution of observed variable $a_{ij}^{u} \in A$ , which is the answer from worker ${u}$ given cell $c_{ij}$, conditioned on parameters $\alpha$, $\beta$ and $\phi$. (Note that worker quality $q_u$  on categorical attributes is replaced by worker quality $\phi_u$ on continuous attributes). 

%, \nikos{why do you mention the tradeoff? can't you just say that you use the standard EM algorithm?} \caihua{yes, it's a standard EM algorighm. Tradeoff means there are other methods can solve it, such as Graphical Model. But considering the run time, I choose EM.}

% \noindent $\bullet$ \textbf{Objective Function. }

% The objective function is 
% \begin{equation}
% \begin{aligned}
% \arg\max  _{\alpha, \beta, \phi} P(A|\alpha, \beta, \phi) = \sum\nolimits_{T} P(A,T|\alpha, \beta, \phi) 
% \end{aligned}
% \end{equation}

%\noindent $\bullet$ \textbf{Expectation Step (E-step). }
\stitle{Expectation Step (E-step)}
 In the E-step, we compute the posterior probabilities of hidden
 variable $T_{ij} \in \mathcal{T}$ given the values of $\alpha$, $\beta$ and $\phi$ and the observed variable $A_{ij} = \{a_{ij}^u\}, u \in U_{ij}$, i.e., the 
current answer set of cell $c_{ij}$.  
%  where $U_{ij}$ is the worker set who answered the cell $c_{ij}$ and $A_{ij}$ is the answer set of the cell $c_{ij}$:
%\vspace{-.5em}
\begin{equation} \label{E_step} %\vspace{-.5em} %\nonumber
\begin{split}
& P(~T_{ij}=z|A_{ij},\alpha_{i}, \beta_{j}, \phi~) \propto \\ 
& \prod\nolimits_{u \in U_{ij}} P(a_{ij}^{u}|T_{ij}=z,\alpha_{i}, \beta_{j}, \phi_{u}) \cdot \text{Prior}(T_{ij}=z).
\end{split}
\end{equation}
% \begin{equation}
% \begin{aligned}
% \label{eq:E_step}
% &P(T_{ij}=z|A_{ij},\alpha_{i}, \beta_{j}, \phi) \\
% \propto &P(T_{ij}=z,A_{ij}|\alpha_{i}, \beta_{j}, \phi) \\
% = &P(A_{ij}|T_{ij}=z,\alpha_{i}, \beta_{j}, \phi) P(T_{ij}=z)\\
% = &\prod_{u \in U_{ij}} P(a_{ij}^{u}|T_{ij}=z,\alpha_{i}, \beta_{j}, \phi_{u}) P(T_{ij}=z)
% \end{aligned}
% \end{equation}
Based on our defined worker model of $P(T_{ij}=z|A_{ij},\alpha_{i}, \beta_{j}, \phi)$ for different datatypes, the distribution is defined as follows. \\
\noindent (1) For cells $c_{ij}$ of continuous type, we regard that $\text{Prior}(T_{ij}=z)$ follows a normal distribution $\mathcal{N}(\mu^{0}_{j},~\phi^{0}_{j})$, and $T_{ij} \sim $ \linebreak $\mathcal{N}({T^{\mu}_{ij},~T^{\phi}_{ij}})$, where $T^{\mu}_{ij}$ and $T^{\phi}_{ij}$ satisfy that 
\begin{equation}  \nonumber
\begin{aligned}
T^{\mu}_{ij} &= \big(\sum\nolimits_{u \in U_{ij}}  \frac{a_{ij}^{u} } {  \alpha_{i}\beta_{j} \phi_{u} } + \frac{ \mu^{0}_{j} } {\phi^{0}_{j}} \big)~T^{\phi}_{ij},\\
T^{\phi}_{ij} &= \big(\sum\nolimits_{u \in U_{ij}}  \frac{1}{ \alpha_{i}\beta_{j} \phi_{u} } + \frac{1}{\phi^{0}_{j}}\big)^{-1}.
\end{aligned}
\end{equation}
%if $\text{Prior}(T_{ij}=z)$ also follows a normal distribution $\mathcal{N}(\mu^{0}_{j},~\phi^{0}_{j})$.\\
% $$
% \begin{aligned}
% %T_{ij} & \sim N(T^{\mu}_{ij}, T^{\phi}_{ij})  \\
% P(T_{ij}=z)  = &\frac{1}{\sqrt{2\pi T^{\phi}_{ij}} }\, e^{-\frac{(z - T^{\mu}_{ij} )^2} {2 T^{\phi}_{ij}} }   (T_{ij} \sim N(T^{\mu}_{ij}, T^{\phi}_{ij}) ) \\
% T^{\mu}_{ij} = &(\sum_{u \in U_{ij}}  \frac{a_{ij}^{u} } { \phi_{u} } + \frac{ \mu^{0}_{j} } {\phi^{0}_{j}} )T^{\phi}_{ij} \\
% T^{\phi}_{ij} = &(\sum_{u \in U_{ij}}  \frac{1}{\phi_{u}} + \frac{1}{\phi^{0}_{j}})^{-1}
% \end{aligned}
% $$
(2) For cells $c_{ij}$ of categorical type, we have 
$$
P(T_{ij}=z) = \frac{ \prod_{u \in U_{ij}} [(q^u_{ij})^{\indfunc{a_{ij}^u=z}} (\frac{1-q^u_{ij}}{|L_j|-1})^ {\indfunc{a_{ij}^u \neq z}} ] }{ \sum_{z\in L_j} \prod_{u \in U_{ij}} [(q^u_{ij})^{\indfunc{a_{ij}^u=z}} (\frac{1-q^u_{ij}}{|L_j|-1})^ {\indfunc{a_{ij}^u \neq z}} ] }, 
$$
where $q^{u}_{ij}$ is defined as $\text{erf}\left( \epsilon / \sqrt{2\alpha_{i}\beta_{j} \phi_{u}}\right)$ and $L_j$ is the label set of column $j$. $\text{Prior}(T_{ij}=z)$ is uniform so it disappears. %We use symbol $T^{p}_{ij}(z)$ to simplify the probability $P(T_{ij}=z)$.

Intuitively, the answer given by high quality worker will be trusted more, i.e., given higher weight. 
To be specific, we estimate the truth distribution $T_{ij}$ by combining
the set $A_{ij}$ of workers' answers for $c_{ij}$. 
(1) $T^\mu_{ij}$ can be regarded as a weighted average of answer $a_{ij}^u$ based on the
quality $\alpha_{i}\beta_{j}\phi_{u}$. $T^\phi_{ij}$ is a normalized term. 
(2) Similarly, $P(T_{ij}=z)$ is a normalized product of
the qualities $q^u_{ij}$ of the workers whose answer $a_{ij}^u$ is $z$.

%\noindent $\bullet$ \textbf{Maximization Step (M-step). }
\stitle{Maximization Step (M-step)}
In the M-step, we find the values of parameters $\alpha$, $\beta$ and $\phi$ that maximize the expectation of the joint log-likelihood of the observed variable $\mathcal{A}$, as shown below: 

\begin{equation} \small \label{M_step}
\begin{aligned}
& Q(\alpha, \beta, \phi)  =  \operatorname{E}_{\mathcal{T}}[ \ln P(\mathcal{A}, \mathcal{T} |\alpha, \beta, \phi)] \\
% = &\sum_{j \in J}\sum_{i \in I} \operatorname{E}_{T_{ij}} [\ln P(a_{ij}, T_{ij} |\alpha_{i}, \beta_{j}, \phi)]  \\
%=&\sum_{j \in J}\sum_{i \in I} \operatorname{E}_{T_{ij}} [\sum_{u \in U_{ij}} \ln P(a_{ij}^{u}|T_{ij},\alpha_{i}, \beta_{j}, \phi_{u})+\ln Prior(T_{ij})] \\
= & \sum_{j}\sum_{i} ~\operatorname{E}_{T_{ij}} \big[\ln \text{Prior}(T_{ij}) + \sum_{u \in U_{ij}} \ln P(a_{ij}^{u}|T_{ij},\alpha_{i}, \beta_{j}, \phi_{u})\big].  
\end{aligned}
\end{equation}

\noindent Formula $\operatorname{E}_{T_{ij}} [\sum_{u \in U_{ij}} \ln P(a_{ij}^{u}|T_{ij},\alpha_{i}, \beta_{j}, \phi_{u})]$ is calculated for the different datatypes, as follows. \\
(1) For cells $c_{ij}$ of continuous type:
$$
\begin{aligned}
%&\operatorname{E}_{T_{ij}} [\sum_{u \in U_{ij}} \ln P(a_{ij}^{u}|T_{ij},\alpha_{i}, \beta_{j}, \phi_{u})]  \\
%& \int_{-\infty}^{+\infty} P(T_{ij}=z) [ \sum_{u \in U_{ij}} \ln(\frac{1}{\sqrt{2\pi \alpha_{i}\beta_{j}\phi_{u}}}\, e^{-\frac{(a_{ij}^{u} - z)^2}{2 \alpha_{i}\beta_{j}\phi_{u}}})]  dz \\
%=&\sum_{u \in U_{ij}} [ - \frac{1}{2}\ln (2\pi \alpha_{i}\beta_{j}\phi_{u}) - \frac{{a_{ij}^u}^2}{2 \alpha_{i}\beta_{j}\phi_{u}} + \frac{2a_{ij}^uT_{ij}^{\mu}}{2 \alpha_{i}\beta_{j}\phi_{u}} - \frac{{T_{ij}^{\mu}}^2+T_{ij}^{\phi}}{2 \alpha_{i}\beta_{j}\phi_{u}}]
\sum\nolimits_{u \in U_{ij}} [ - \frac{1}{2}\ln (2 \pi \alpha_{i}\beta_{j}\phi_{u}) - \frac{({a_{ij}^u} -T_{ij}^{\mu})^2 +T_{ij}^{\phi}}  {2 \alpha_{i}\beta_{j}\phi_{u}} ].
\end{aligned}
$$
(2) For cells $c_{ij}$ of categorical type:
\begin{equation} \nonumber
\begin{split}
%&\operatorname{E}_{T_{ij}} [\sum_{u \in U_{ij}} \ln P(a_{ij}^{u}|T_{ij},\alpha_{i}, \beta_{j}, \phi_{u})]  \\
%& \sum_{z \in L_j }  P(T_{ij}=z)[ \sum_{u \in U_{ij}} \ln (q_u)^{\indfunc{a_{ij}^u=z}} (\frac{1-q_u}{|L_j|-1})^ {\indfunc{a_{ij}^u \neq z}}] \\
\sum\nolimits_{z \in L_j } P(T_{ij}=z) \cdot \sum\nolimits_{u \in U_{ij}} \big(  & {\indfunc{a_{ij}^u=z}} \ln \text{erf}( \frac{\epsilon}{\sqrt{2\alpha_{i}\beta_{j} \phi_{u}}})  \\
+ & {\indfunc{a_{ij}^u \neq z}} \ln \frac{1-\text{erf}( \frac{\epsilon}{\sqrt{2\alpha_{i}\beta_{j} \phi_{u}}})}{|L_j|-1}\big).
\end{split}
\end{equation}

We apply gradient descent to find the values of $\alpha$, $\beta$ and $\phi$ that locally maximize $Q(\alpha, \beta, \phi)$.

Intuitively, a worker will be of high quality if his/her answers are close to the estimated truth. 
Thus, we compute a value $\phi_u$ that maximizes the
expectation of the log-likelihood of worker $u$'s answers
$a_{**}^u$. Similarly, we also find an $\alpha_i$ (resp. $\beta_j$)
that maximizes the expectation of the log-likelihood of answers
$a_{i*}^*$ in row $i$ (resp. $a_{*j}^*$ in column $j$). 
%However, since
%the answers are overlap, $\alpha_i$, $\beta_j$ and $\phi_u$ cannot be computed independently and we estimate them by gradient descent.

%\noindent $\bullet$ \textbf{Algorithm. }  
\stitle{Algorithm}
By combining the two steps above, we can iteratively update the parameters until convergence. Each $T_{ij}$ is initialized 
%(e.g., by 
by following the distribution in $\text{Prior}(T_{ij})$.
%), we run an iterative approach.
At each iteration, the M-step applies gradient descent to find $\alpha_i$, $\beta_j$ and $\phi_u$ by maximizing Equation~\ref{M_step} and the E-step applies Equation~\ref{E_step}. 
We identify convergence if the differences between the parameter values in subsequent iterations are below a threshold (e.g., $10^{-5}$). 

Finally we estimate the truth $\widehat{T}_{ij}$ of each cell $c_{ij}$ as:
$$
\widehat{T}_{ij} = 
\begin{cases}
T^{\mu}_{ij}  &,~ c_{ij} \text{ is continuous}, \\
\arg \max_{z\in {L_j}} ~P(T_{ij}=z) &,~ c_{ij} \text{ is categorical}.
\end{cases} 
$$
Algorithm \ref{alg:inference} shows the detailed process of inference.

%\stitle{Preprocessing}
%Since different continuous attributes may have different scales, we transform them to their z-scores for the following inference process. Specifically, we compute median $m_{ij}$ of answers in each cell and standard deviation $\sigma_j$ of each column. For each answer $a_{ij}^u$, we normalize it as 
%$$\hat{a_{ij}^u} = \frac{ a_{ij}^u - m_{ij} }{\sigma_j}.$$
%After inferring the truth $\hat{T_{ij}}$ for a continuous cell
%$c_{ij}$ based on $\hat{a_{ij}^u}$, we obtain the true value in the
%original scale by $\hat{T_{ij}} \cdot \sigma_j + m_{ij}.$

\stitle{Time Complexity} 
In the E-step, computing hidden variable $T_{ij}$ for a continuous
cell $c_{ij}$ requires looping through the observed variable $A_{ij} =
\{a_{ij}^u\}$, hence the complexity is $\mathcal{O}(|A_{ij}|)$. 
For a categorical cell $c_{ij}$, we need to additionally loop through
the possible answers, thus the cost becomes
$\mathcal{O}(l \cdot |A_{ij}|)$ where
 $l= \text{max}_{j}(|L_j|)$.
 % so it takes $\mathcal{O}(l \cdot |A_{ij}|)$. 
The total cost of the E-step is therefore $\mathcal{O}(l \cdot
|\mathcal{A}|)$, where $\mathcal{A}$ is the set of all obtained
answers.
In the M-step, to compute $Q(\alpha, \beta, \phi)$, we need to loop
for each cell and workers who answered this cell. 
The number of loops is the same as $|\mathcal{A}|$. 
Since we use gradient descent, we need to also compute the gradient of
each parameter which takes $\mathcal{O}(l \cdot |\mathcal{A}|)$. 
If gradient descent takes $v$ iterations to converge, this step takes
$\mathcal{O}(v l \cdot |\mathcal{A}|)$ time in total. Assuming that
the algorithm
needs $w$ iterations to converge, the total time complexity is $\mathcal{O}(w v l \cdot |\mathcal{A}|)$. In practice, $l$ is constant, and $v$ and $w$ are smaller than 20, thus the time complexity is linear to the number of answers.

\begin{algorithm}[!ht]\small
   \caption{\emph{Truth Inference Method}}  \label{alg:inference}
\KwIn  {workers' answers $a_{ij}^u \in \mathcal{A}$, prior distribution of truth $\text{Prior}(T_{ij})$ }
\KwOut {truth distribution $T_{ij} \in \mathcal{T}$, worker's quality $\phi_u$, difficulty of row $\alpha_i$ and column $\beta_j$ }   

   {Initialize $T_{ij}$ using $\text{Prior}(T_{ij})$ }\\ \label{lines:initialize}
   \While { \textbf{true} } {

   \vspace{.3em}
   {// \emph{Step 1: Estimate Worker Quality and Cell Difficulty}}\\
   {Compute $\alpha_i$, $\beta_j$ and $\phi_u$ maximizing Eq.~\ref{M_step};}

   \vspace{.3em}
   {// \emph{Step 2: Infer the Truth}}\\
   \For{$1\le i\le N$}{ 
   	\For{$1\le j\le M$}{
   	      { Obtain $T_{ij}$ by Eq.~\ref{E_step}; }\\
   	 }
   }
	\vspace{.3em}
   {// \emph{Check for Convergence}} \\ 
   \If {Converged}
          {\textbf{break};}
   
   }
	\Return {$T_{ij}$, $\alpha_i$, $\beta_j$ and $\phi_u$;}\\

\end{algorithm}

%!TEX root = main.tex
%\newpage
\section{Online Task Assignment}\label{sec:Taskassignment}

In this section, we discuss how we select tasks for a worker $u$. 
Section \ref{sec:inherentIG} defines an {\it inherent information gain} function to measure the utility of assigning a task to the worker, which can handle both categorical and continuous data.
The function considers the quality of the worker, the need to obtain more answers for the task, and the task's difficulty. Intuitively, we prefer to assign tasks whose gain of information will be improved the most if the incoming worker answers them. 
% based on the quality of the worker. 
%Information gain also evaluates whether a task is really beneficial to be assigned. 
In Section \ref{sec:task:corr}, we extend this to a 
{\it structure-aware information gain} function, which also considers the correlations in the qualities of answers given by the same worker to different cells of the same row. 
%attributes into {\it inherent information gain} function. Thus, it estimates utility of a task more accuratly. 
%Section \ref{sec:task:K} discusses the assignment of $K>1$ tasks simultaneously. 
Section \ref{sec:task:K} discusses the assignment of multiple tasks to $u$.

\subsection{Inherent Information Gain} \label{sec:inherentIG}

%As we have both categorical and continuous tasks in the structured data, 
We need a uniform measure for the {\it utility} (or benefit) of assigning a task (either categorical or continuous) to a worker $u$ with quality $q_u$. 
For this purpose we define an {\it inherent information gain} function, following the steps below.
%We introduce a measure of utility, which we name {\it inherent information gain}. For categorical and continuous cell, we discuss them, respectively. 

\noindent (1) For a categorical cell $c_{ij}$, the distribution of truth $T_{ij}$ has been computed by $P(T_{ij}=z)$ in Equation \ref{E_step}, which is the probability that label $z$ is correct. Thus, Shannon Entropy \cite{entropy}, a well-studied measure, can be used to define the uncertainty of task $c_{ij}$:  
$$
H_s(T_{ij}) = -\sum\nolimits_{z \in L_j } P(T_{ij}=z) \ln{P(T_{ij}=z)}.
$$

\noindent (2) For a continuous cell $c_{ij}$, note that for a continuous distribution, the Differential Entropy \cite{differential_entropy} is defined as:
$$ -\int_\mathbb{X} f(x)\ln f(x)\,dx , $$
where $f(x)$ is a probability distribution. Recall that we also define the distribution of truth $T_{ij} \sim \mathcal{N}(T_{ij}^{\mu}, T_{ij}^{\phi})$ of a continuous cell $c_{ij}$ in Equation \ref{E_step}, so its Differential Entropy can be computed as:
$$
H_d(T_{ij}) = \frac{1}{2}\,{\ln{\left(2\,\pi\,e\, T^{\phi}_{ij} \right)}}.
$$

Given the above, we define a {\em uniform entropy} for a task $c_{ij}$ as:
$$
H(T_{ij}) = 
\begin{cases}
H_d(T_{ij}),&~ \text{if } c_{ij} \text{ is continuous}, \\
H_s(T_{ij}),&~ \text{if } c_{ij} \text{ is categorical}.
\end{cases} 
$$

A straightforward approach for task assignment to a worker $u$ is to select the task $c_{ij}$ with the largest uniform entropy. However, this is problematic, as Differential Entropy and Shannon Entropy are not comparable; hence, task assignments may be biased toward one datatype. For example, as pointed out in \cite{differential_entropy}, Differential Entropy can be negative while Shannon entropy is always non-negative. 
%If we use entropy, task assignments may biased toward one datatype. 

Alternatively, we use Delta Entropy to measure the information gain. Suppose $\mathcal{A}_C$ is the current set of answers we have collected,
%\nikos{why use a different symbol? In section 3 and in the table of notations, you use $\mathcal{A}$}\caihua{I use $\mathcal{A}$ as a set of answers. $\mathcal{A}_C$ means current answers. }, 
we can obtain the estimated truth distribution (denoted as $T_{ij,\mathcal{A}_C}$) for each task $c_{ij}$ by the truth inference method presented in Section~\ref{sec:TruthInference}. Specifically, for an incoming worker $u$, we define the inherent information gain of assigning task $c_{ij}$ to her as:
% and gives an answer $a^u_{ij}$ for task $c_{ij}$,
%is assigned a task $c_{ij}$. Suppose his answer is $a^u_{ij}$,
% we can update the estimated truth distribution as $T_{ij,\mathcal{A}_C + a^u_{ij}}$.
%Then, the inherent information gain for this new answering is defined as:
\begin{equation} \label{eq:ig}
IG_q(c_{ij}) =  H(T_{ij,\mathcal{A}_C}) - \operatorname{E}_{a_{ij}^{u}}[ H(T_{ij,\mathcal{A}_C \cup \{a^u_{ij}\}})],
\end{equation}
 where $T_{ij,\mathcal{A}_C \cup \{a^u_{ij}\}}$ is the updated distribution of the estimated truth for task $c_{ij}$ after receiving a new answer $a_{ij}^u$ from worker $u$. 
 
By using the inherent information gain measure defined in Equation \ref{eq:ig}, we alleviate the problem that the domains of the two entropy types are different.
%, making delta entropies comparable.
%\nikos{There are still problems here. First, if the scales in the domains of the two entropies are different, then one would be favored over the other. So, the delta entropy may not solve the problem of potentially different scales. This seems to be solved by the discussion below, still the support is based on a hypothesis. Second, how do you know $\widehat{T_{ij}}$? This is posterior knowledge, but you need it to decide $c_{ij}$. I mean do you need a workers real answer on the cell in order to compute the delta entropy and decide if you are going to put this cell in a HIT? This is not reasonable.} \caihua{different scales is very difficult to solve and I have no idea now. So I said that they can be compared in aspect of definition.}
If we discretize the range of a continuous random variable $X$ using bins of width $\Delta$, we can compute the Shannon entropy for this new discretized random variable $X^\Delta$, and we have the following formula if $X$'s pdf is Riemann integrable: 
$$
H_s(X^\Delta)+ \ln\Delta \to H_d(X), ~as~\Delta \to 0.
$$
Hence, if $\Delta$ is small, $H_d(X_1) - H_d(X_2) \approx H_s(X_1^\Delta) -H_s(X_2^\Delta)$, which means that the subtraction of differential entropies can be transformed into subtraction of Shannon entropies. As a result, for cells of different types, $IG(c_{ij})$ is comparable. Algorithm \ref{alg:assignment} shows the detailed process of assignment.

\vspace{0.4em}
\noindent {\bf Computing the distribution of $\operatorname{E}_{a_{ij}^{u}}[ H(T_{ij,\mathcal{A}_C \cup \{a^u_{ij}\}})]$.} The distribution of an answer $a_{ij}^u$ follows the worker model in Equations \ref{eq:contap} and \ref{eq:cateap} for continuous and categorical tasks, respectively.
% If the worker $u$ is a new one and does not have any history, then his quality $q_u$ is assumed to be the median of all known workers.
For a categorical task $c_{ij}$, the domain of $a_{ij}^{u}$ is a finite label set, so we use all possible values $a_{ij}^{u}$ to obtain $T_{ij,\mathcal{A}_C \cup \{a_{ij}^u\}}$ using the inference method described in Section \ref{sec:TruthInference}.
For a continuous task, since the the domain of $a_{ij}^{u}$ is $\mathcal{R}$, we apply sampling to approximate the value of $T_{ij,\mathcal{A}_C \cup \{a_{ij}^u\}}$.
However, it is quite expensive to run the inference method for each possible answer.
%time-consuming for one possible answer to run the inference method again. 
Because one more possible answer is quiet small compared with current set of answers, we accelerate by updating the parameters related to this answer mostly and maintaining other parameters. Thus, for a new answer $a_{ij}^u$, we update truth distribution $T_{ij}$, and the qualities of workers who have answered task $c_{ij}$.

\vspace{0.4em}
\noindent {\bf Time Complexity.} To compute the benefit for each task $c_{ij}$ (Equation \ref{eq:ig}), we should first iterate through the possible answers given by the incoming worker and compute a new distribution of truth $T_{ij}$. The number of possible answers for a categorical task $c_{ij}$ is $|L_j|$ and for a continuous task is the fixed sampling number $s_\text{cont}$. %Because we approximate the inference method, it takes $\mathcal{O}(v l \cdot |\mathcal{A}|)$. Let $s=\text{max}(\text{max}_{j}(|L_j|), s_\text{cont})$; the total cost of considering one task for a certain worker is $\mathcal{O}(svl \cdot |\mathcal{A}|)$. Then, computing the information gains of all tasks takes $\mathcal{O}(NMsvl \cdot |\mathcal{A}|)$. 
Because we approximate the inference method, it only takes $\mathcal{O}(l \cdot |P|)$ where $P$ is the set of parameters we need to update. Let $s=\text{max}(\text{max}_{j}(|L_j|), s_\text{cont})$; the total cost of considering one task for a certain worker is $\mathcal{O}(sl \cdot |P|)$. Then, computing the information gains of all tasks takes $\mathcal{O}(NMsl \cdot |P|)$. Since $P$ includes the truth distribution $T_{ij}$ and the qualities of workers who have answered task $c_{ij}$, $P$ mainly depends on the average answers per task. Thus, $\mathcal{O}(NMsl\cdot |P|) \approx \mathcal{O}(sl \cdot |\mathcal{A}|)$.

Parallel or distributed computation can be used to accelerate task assignment, as the consideration of the different tasks are independent. For example, we separate different tasks and different possible answers to different machines or processes and compute the corresponding information gains in parallel without the need of data communication.

\begin{algorithm}[!t]\small
   \caption{\emph{Online Task Assignment Method}}  \label{alg:assignment}
\KwIn  {Budget $B$}
\KwOut {truth distribution $T_{ij} \in \mathcal{T}$ }   

   {Initialize each task with several answers from workers }\\ \label{lines:initialize}

   \While { \textbf{Budget $B$ is not exhausted} } {

   \vspace{.4em}
   {// \emph{Step 1: Analyze current situation}}\\
   {Run truth inference to obtain $T_{ij}$, $\alpha_i$, $\beta_j$ and $\phi_u$}
   %{Compute correlation $P(e_{j}|e_{k})$ and correlation coefficient $W_{jk}$}

   \vspace{.5em}
   {// \emph{Step 2: Find task $c^*$ with highest benefit for incoming worker $u$}}\\
   
   \For{$1\le i\le N$}{ 
   	\For{$1\le j\le M$}{
   		  %{$c^* = \text{arg max}_{c_{ij}}{IG(c_{ij})}$}
   	      {Compute information gain $IG(c_{ij})$ by Eq.\ref{eq:ig}}\\
   	      \If {$IG(c_{ij}) >  IG(c^*)$ or $c^*$ \text{is not defined}}
          		{$c^* = c_{ij}$}
   	 }
   }

   \vspace{.4em}
   {// \emph{Step 3: Collect answers}}\\
   {Publish task $c^*$ and collect worker $u$'s answer}
   }
   
    {Run truth inference to obtain the final $T_{ij}$} \\
	\Return {$T_{ij}$}\\

\end{algorithm}

\subsection{Structure-Aware Information Gain}\label{sec:task:corr}

%\reminder{correlation is not a good word. because reviewers will ask how to get the correlation!}

The task assignment approach based on inherent information gain, described in Section \ref{sec:inherentIG}, does not utilize the structural information of $C$.
We now propose a structure-aware task assignment method. The basic idea is to estimate correlation, i.e., the conditional distribution of the error on a task $c_{ij}$, given the errors on other tasks $c_{ik}$ in the same row. For this, we consider the answer history of all the workers and then use the conditional distribution to obtain a better estimation of the target worker $u$'s error on task $c_{ij}$.
%\footnote{The reason that we do not estimate the probability among three or more columns is data sparsity. The known answers are typically not too many, so the estimated probability will be inaccurate.} 

We first define $e_j$ as a random variable for the error of all the answers $a_{*j}^*$ on attribute $j$, and we view the error of each answer $e_{ij}^u $ as a sample of $e_j$. For a continuous attribute, $e_{ij}^u= a_{ij}^u - \widehat{T}_{ij}$, while for a categorical attribute, $e_{ij}^u$ is simply 0 or 1, considering whether $a_{ij}^u$ equals $\widehat{T}_{ij}$.

Suppose worker $u$ has answered task $c_{ik}$ before; then,
% we have:
$$P\left(e_{ij}^u|e_{ik}^u\right) = P\left(e_{j} | e_{k}=e_{ik}^u \right),$$
where $P(e_j|e_k)$ is the correlation between attributes $j$ and $k$. We estimate $P(e_j|e_k)$ with a maximum likelihood method considering all the answers we have collected. The computation of $P(e_j|e_k)$ is discussed later. 

When worker $u$ has answered multiple tasks on row $i$, we need to consider all the observed errors for row $i$. However, it is not practical to estimate the conditional distribution, given errors from multiple attributes, due to data sparsity. Hence, we consider a linear combination of the correlations. Formally, we have:
\begin{equation}
\label{eq:correlated_quality}
P\left( e_{ij}^u|E_{i}^u \right) = \frac{\sum_{c_{ik} \in L_i^u }  W_{jk} \cdot P \left(e_{j} | e_{k}=e_{ik}^u \right)}
 {\sum_{c_{ik} \in L_i^u }  W_{jk} }, 
\end{equation}
where $L_i^u$ is the set of tasks which worker $u$ has answered on row $i$ and the observed error set $E_{i}^u=\{e_{ik}^u|c_{ik} \in L_i^u\}$. $W_{jk}$ is the correlation coefficient between attribute $j$ and $k$:
%, defined as:
\begin{equation}
W_{jk} =\frac{(M_{j} - \overline{M}_j)(M_{k} - \overline{M}_k)}{\sqrt{(M_{j} - \overline{M}_j)^2} \sqrt{(M_{k} - \overline{M}_k)^2}}, 
\end{equation}
where $M_j$ ($M_k$) is the error vector on attribute $j$ ($k$) and each element in the vector, i.e., $e^*_{*j}$ ($e^*_{*k}$) is defined as above. $\overline{M}_j$ ($\overline{M}_k$) is also a vector, where each element is the mean of vector $M_j$ ($M_k$).

After obtaining the conditional distribution of $e_{ij}^u$, we also obtain a more accurate distribution of answer $a^u_{ij}$. 
Then, we calculate the structure-aware information gain $IG_{c}(c_{ij})$ using Equation \ref{eq:ig}.

\noindent {\bf Computing the Correlation $P(e_{j}|e_{k})$.} Correlation is defined as the conditional probability between two columns $j$ and $k$ which can also be estimated from known answers. Since we have categorical and continuous columns, we have four cases in total as shown in Table \ref{table:fourcases}. We calculate the joint probability and the marginal probability, and obtain the conditional probability by:
$$
P(e_{j}|e_{k}) = \frac{P(e_{j},e_{k})}{P(e_{k})}.
$$
\noindent \textbf{(1) Marginal distribution.} As shown in Table \ref{table:twocases}, we introduce the marginal probability $P(e_{j})$. Following the definition above, a categorical column is regard as a Bernoulli distribution so we estimate $\psi^p_{j}=P(e_{j})$. Similarly, a continuous column is a Normal distribution, so we estimate $\psi^\mu_{j}$ as mean and $\psi^\phi_{j}$ as variance.

\begin{table}[]
\centering 
\caption{Marginal Distribution $P(e_{j})$} \scriptsize
\label{table:twocases}
\begin{tabular}{|@{~}c@{~}|@{~}c@{~}|@{~}c@{~}|}
\hline
\textbf{Type $j$}       & \textbf{Distribution}           & \textbf{Estimated Parameter(s)} \\ \hline \hline
Categorical & Bernoulli,  $\mathcal{B}(1,\psi^p_{j})$ & $\psi^p_{j}$   \\ \hline
Continuous  & Normal,   $\mathcal{N}(\psi^\mu_{j}, \psi^\phi_{j})$  &  $\psi^\mu_{j}$, $\psi^\phi_{j}$   \\ \hline
\end{tabular}
\end{table}

\begin{table}[]
\centering
\caption{Conditional Distribution $P(e_{j}|e_{k})$} \scriptsize
\label{table:fourcases}
\begin{tabular}{|@{~}c@{~}|@{~}c@{~}|@{~}c@{~}|@{~}c@{~}|}
\hline
\textbf{Type $j$}   & \textbf{Type $k$}     & \textbf{Distribution}           & \textbf{Estimated Parameter(s)}\\ \hline \hline
Categorical & Categorical & Bernoulli &  $\psi^{p_r}_{jk}, \psi^{p_w}_{jk}$  \\ \hline 
Continuous  & Continuous & Normal     &  $\psi^{\vec{\mu}}_{jk}, \psi^{\Sigma}_{jk}$   \\ \hline 
Categorical & Continuous & Bernoulli   &  $\psi^{\mu_r}_{kj}, \psi^{\phi_r}_{kj}, \psi^{\mu_w}_{kj}, \psi^{\phi_w}_{kj}$ \\ \hline 
Continuous & Categorical & Normal  &   $\psi^{\mu_r}_{jk}, \psi^{\phi_r}_{jk}, \psi^{\mu_w}_{jk}, \psi^{\phi_w}_{jk}$   \\ \hline 
\end{tabular}
\end{table}

%Based on the known answers, it's easy to estimate.
% \noindent \textbf{Joint distribution.} We define the joint distribution $P(e^u_{ij},e^u_{ik})$, for any row and worker $(i,u)$. We have three situations based on the different types of columns $j$ and $k$.\\
% 1) \textbf{both categorical:} Because Bernoulli distribution has two situations, right or wrong, $P(e^u_{ij},e^u_{ik})$ has four situations and estimate these probabilities $P(e^u_{ij}=0,e^u_{ik}=0)$, $P(e^u_{ij}=1,e^u_{ik}=0)$, $P(e^u_{ij}=0,e^u_{ik}=1)$ and $P(e^u_{ij}=1,e^u_{ik}=1)$.  \\
% 2) \textbf{both continuous:} Because both columns follow normal distributions, $P(e^u_{ij},e^u_{ik})$ is a bivariate normal distribution and we estimate the $\psi^{\vec{\mu}}_{jk}$ as mean and $\psi^{\Sigma}_{jk}$ as sigma. \\
% 3) \textbf{one is categorical and another is continuous:} w.l.o.g, let $j$ be categorical and $k$ be continuous. We approximately assume that the conditional distributions $P(e^u_{ik}|e^u_{ij}=0)$ and $P(e^u_{ik}|e^u_{ij}=1)$ obey a normal distribution. To model $P(e^u_{ij},e^u_{ik})$, we need firstly to estimate four parameters, $\psi^{\mu_{r}}_{jk}$ as mean and $\psi^{\phi_r}_{jk}$ as variance given $e^u_{ij}=0$ and $\psi^{\mu_{w}}_{j}$ and $\psi^{\phi_w}_{j}$ given $e^u_{ij}=1$. Based on multiplying $P(e^u_{ij}=0)$ or $P(e^u_{ij}=1)$,we obtain $P(e^u_{ik}, e^u_{ij}=0)$ and $P(e^u_{ik}, e^u_{ij}=1)$ separately.

\noindent \textbf{(2) Conditional distribution.} We now explain how to obtain the conditional probability $P(e_{j}|e_{k})$. W.l.o.g., we assume we already know worker $u$'s answer is $a_{ik}^u$ so we know $e_{ik}^u$; then, we have the four cases shown in Table \ref{table:fourcases}. For each of these cases, we use maximal likelihood method to estimate the corresponding parameters. We elaborate on the four cases below:
%list the corresponding parameters as follows: 

\vspace{0.4em}
\textbf{(a) both categorical:} we need to estimate two parameters: $\psi^{p_r}_{jk} = P(e_{j} | e_{k}= e_{ik}^{u} = 0)$ and $\psi^{p_w}_{jk} = P(e_{j} | e_{k}= e_{ik}^{u}= 1)$ separately. 

\vspace{0.4em}
\textbf{(b) both continuous:} Because both columns follow normal distributions, the joint distribution $P(e_{j},e_{k})$ is a bivariate normal distribution and we estimate $\psi^{\vec{\mu}}_{jk}$ as the mean vector and $\psi^{\Sigma}_{jk}$ as the covariance matrix. Assume that $\psi^{\vec{\mu}}_{jk} = \begin{pmatrix}
\mu_j \\
\mu_k
\end{pmatrix}
$
, $\psi^{\Sigma}_{jk} =
\begin{pmatrix}
    \sigma_j^2 & \rho\sigma_j\sigma_k \\
     \rho\sigma_j\sigma_k & \sigma_k^2
\end{pmatrix}
$.
The conditional distribution $P(e_{j}|e_{k}=e_{ik}^{u})$ is also a normal distribution $\mathcal{N}(\mu_j+\frac{\sigma_j}{\sigma_k}\rho( x - \mu_k) ,\,  (1-\rho^2)\sigma_j^2 )$.

\vspace{0.4em}
\textbf{(c) column $k$ is categorical and column $j$ is continuous:} We assume that the conditional distributions $P(e_{j}|e_{k}=0)$ and $P(e_{j}|e_{k}=1)$ obey normal distributions. We obtain $\mathcal{N}(\psi^{\mu_r}_{jk}, \psi^{\phi_r}_{jk})$ when $e^u_{ij}=0$ and $\mathcal{N}(\psi^{\mu_w}_{jk}, \psi^{\phi_w}_{jk})$ when $e^u_{ij}=1$. 

\vspace{0.4em}
\textbf{(d) column $j$ is categorical and column $k$ is continuous:}  
Based on the same assumptions as in case c), we can estimate that $P(e_k|e_j=0)$ follows $\mathcal{N}(\psi^{\mu_w}_{kj}, \psi^{\phi_w}_{kj})$ and $P(e_k|e_j=1)$ follows $\mathcal{N}(\psi^{\mu_r}_{kj}, \psi^{\phi_r}_{kj})$. Because we also know $P(e_k=e^u_{ik})$, $P(e_j=0)$ and $P(e_j=1)$, we calculate conditional distributions $P(e_{j}=0|e_{k}=e^u_{ik})$ 
%=  \frac{P(e_{k}=e^u_{ik}| e_{j}=0)P(e_{j}=0)}{P(e_{k}=e^u_{ik})}$ 
and $P(e_{j}=1|e_{k}=e^u_{ik})$ accordingly.

\vspace{0.4em}
\noindent {\bf Time Complexity.} To compute the correlation $P(e_{j}|e_{k})$, we should iterate through each column and calculate the corresponding conditional distribution. Because there are $M$ columns, the total cost is 
% and have to perform the we need to loop all the answers to compute the correlation, it takes 
$\mathcal{O}(M \cdot |\mathcal{A}|)$. The same time is needed to calculate the correlation coefficient $W_{jk}$. The cost of computing the benefit of each task is the same as that of computing the Inherent Information Gain, which is discussed before. In total, the cost is $\mathcal{O}((M+sl) \cdot |\mathcal{A}| )$. %Note that the time for computing correlations is negligible compared to the time for required for computing the information gain.

%, which is the same as inherent information gain.

%The inherent information gain $IG_q$ and correlation-aware information gain $IG_c$ consider two different aspects: $IG_q$ is based on worker's quality $q_u$, while $IG_{c}$ is based on the correlation among columns. 
%Using a linear combination, we further consider a hybrid information gain as: 
%\begin{equation}\label{eq:linear}
%IG_h(c_{ij}) = \kappa IG_{c}(c_{ij}) + (1-\kappa) IG_q(c_{ij})
%\end{equation}
%where $\kappa \in [0,1]$ is a parameter to balance two sources of gain. $\kappa$ can be given by the requester or set by the system according to historical performance. In section \ref{exp:test_kappa}, we show the performance with different $\kappa$.
%
%\nikos{unclear how to determine $\kappa$, it would be nice to have an approach that finds the best $\kappa$ according to the current situation (knowledge about the worker or knowledge about the correlation between columns).}

\subsection{Assigning Multiple Tasks to Workers}\label{sec:task:K}

So far we focused on how to select one task to assign to the incoming worker. This does not restrict the applicability of our approach in the case that multiple tasks should be determined and given to the worker as a batch (e.g., as in a HIT on AMT~\cite{AMT}).
%, when a worker comes, often a set of $K$ tasks are dynamically selected and batched in a HIT (Human Intelligence Task) to assign to the coming worker. 
%We next address how to select $K$ tasks (denoted as $D=\{c_{i_1j_1}, c_{i_2j_2}, \cdots, $ $c_{i_Kj_K} \} $) to the coming worker. 
Suppose that the worker is to be assigned a set $D=\{c_{i_1j_1}, c_{i_2j_2}, \cdots, $ $c_{i_Kj_K} \} $ of $K$ tasks. From the set 
$\mathcal{A_{D}} = \{a_{i_1j_1}^{u}, a_{i_2j_2}^{u}, \cdots, a_{i_kj_k}^{u}\}$ 
of  estimated answers to the tasks by the worker, 
we can update the distribution of the estimated truth $T_{ij,\mathcal{A}_C \cup \mathcal{A_{D}} }$ for each task $c_{ij} \in D$. Then, we can calculate the information gain for $D$ as:
\begin{equation}
\label{eq:task:K}
IG(D) =  \sum\nolimits_{c_{ij} \in D} \big(H(T_{ij,\mathcal{A}_C}) - \operatorname{E}_{\mathcal{A_{D}} } [ H(T_{ij,\mathcal{A}_C \cup \mathcal{A}_D})]\big). 
\end{equation}

Because the search space of $D$ is $\binom{N \cdot M}{K}$, 
%and $ \operatorname{E}_{\mathcal{A_{D}} } [ H(T_{ij,\mathcal{A}_C+\mathcal{A}_D})] $ is hard to simply
finding $K$ tasks which maximize $IG(D)$ is expensive. %This problem can be proved as NP-hard. 
To alleviate the cost, we can apply a greedy approach that iteratively selects the top-$K$ tasks with the largest $IG(c_{ij})$.

\section{Experiments}\label{sec:experiments}
We now present the experimental results. We discuss the experiment datasets in Section~\ref{subsec:data}. In Sections~\ref{subsec:TI} and \ref{subsec:TA}, we compare different crowdsourcing solutions in terms of truth inference and task assignment respectively. We perform case studies in Section~\ref{subsec:case}. Results on synthetic datasets are shown in Section~\ref{sec:exp:synth}.

We have implemented a prototype of T-Crowd and other crowdsourcing solutions in Python 2.7, on a Ubuntu server with 8-core Intel(R) Core(TM) i7-3770 CPU @ 1.60GHz cores and 16 GB memory.

%We compare T-Crowd with state-of-the-art crowdsourcing solutions. 
%with respect to
%its effectiveness in truth inference and task assignment. 

\subsection{Datasets}
\label{subsec:data}

% \begin{table}[]
% \centering
% \caption{Statistics of Real-world Datasets} \small
% \label{table:dataset_setting}
% \begin{tabular}{|@{~}c@{~}|@{~}c@{~}|@{~}c@{~}|@{~}c@{~}|@{~}c@{~}|@{~}c@{~}|} 
% \hline
% %\hline\begin{tabular}[c]{@{}c@{}}\#answers\\  per task\end{tabular}
% \textbf{dataset}         & \textbf{\#rows} & \textbf{\#columns} & \textbf{\#cells} & \textbf{\#answers}  & \textbf{\#ans. per task} \\ \hline \hline
% Celebrity   & 174    & 7         & 1218    &6090       & 5 \\ \hline                                                                    
% Restaurant  & 203    & 5         & 1015    &4060       & 4 \\ \hline                                                                   
% Emotion     & 100    & 7         & 700     &7000      & 10 \\ \hline                                                                   
% \end{tabular}
% \end{table}

\begin{table}[]
\centering
\caption{Statistics of Real-world Datasets} \small
\label{table:dataset_setting}
\begin{tabular}{|@{~}c@{~}|@{~}c@{~}|@{~}c@{~}|@{~}c@{~}|@{~}c@{~}|} 
\hline
%\hline\begin{tabular}[c]{@{}c@{}}\#answers\\  per task\end{tabular}
\textbf{Dataset}         & \textbf{\#Rows} & \textbf{\#Columns} & \textbf{\#Cells}  & \textbf{\#Ans. per Task} \\ \hline \hline
Celebrity   & 174    & 7         & 1218          & 5 \\ \hline                                                                    
Restaurant  & 203    & 5         & 1015          & 4 \\ \hline                                                                   
Emotion     & 100    & 7         & 700           & 10 \\ \hline                                                                   
\end{tabular}
\end{table}

We use three real datasets to perform the experiments. Their statistics are shown in Table \ref{table:dataset_setting}.
%We have crowdsourced three real databases that have ground truth (Table \ref{table:dataset_setting}), as detailed below.
%Celebrity and Restaurant are collected via  
%We used the popular crowdsourcing platform 
% Amazon Mechanical Turk
% (AMT~\cite{AMT}).
%  %to collect two real datasets (Celebrity Information Collection and Restaurant Review Analysis).
%  We make sure that each task (i.e., cell) is answered 5 times for Celebrity and 4 times for Restaurant, with each answer costing 0.2 cents approximately.
% % \nikos{do you mean at least 5 times? how can you guarantee good quality for difficult cells that may need additional answers? in general, should some cells get more answers than others?}\zhipeng{no, it is exactly 5 times. The public dataset also have a fix number of answers for each task (they have 10. See the Emotion data)}
%  Emotion is a public dataset \cite{Emotiondataset}, where each task is answered 10 times.
 % And we collect them by row, where each row costs \$0.01.
%We also used a publicly available dataset (Emotion Data from \cite{Emotiondataset}), which each task is answered 10 times and also collect by row. \\

%\noindent $\bullet$ Celebrity: 
\stitle{Celebrity~\cite{celedataset}}
This dataset contains the information of celebrities. 
Given a celebrity's picture, workers are requested to provide the following attribute values:
name, age, height, nationality, ethnicity, notability, and facial expression of the celebrity in the picture. While name, nationality, and ethnicity are categorical, age, height, notability, and facial are continuous.
The ground truth for name and age are obtained from \cite{celedataset}. 
%It provides pictures and ground truth of name and age accordingly. 
We label the true values of other attributes manually. %We perform experiments on AMT~\cite{AMT}, where each task is answered 4 times by different workers. 
%\reynold{Is this ``manual label'' correct?}

%\noindent $\bullet$ Restaurant: 
\stitle{Restaurant~\cite{ABSAdataset}}
This dataset contains
% \reynold{review?} 
information about restaurants. 
Given a review about a certain
restaurant, workers are asked to specify the aspect (e.g., food or location), attribute (e.g., price or style),
%\reynold{The attribute is called ``attribute'' itself, which is quite confusing. How about change it to ``property''?}
and sentiment (e.g., negative or positive) of the review. 
They are asked to identify the target (i.e., the restaurant referred by the review) by its starting and end position in the text. Here, aspect, attribute, and sentiment are categorical; the starting and end positions are continuous.
We obtain the reviews and ground truth of the restaurants from %Semeval-2014 task 4 : Aspect-Based Sentiment Analysis 
\cite{ABSAdataset}.
%\nikos{remove Citeseer from that reference. In general make a careful pass over the references and clean them.}\caihua{solved}
% of restaurant reviews.
%\nikos{it is not clear to me what attribute means. also, what does 'target of review' mean?} \zhipeng{target is the entity that the worker is giving review to. For example, if the review is about the taste of the burger, the target is burger, and the worker is required to identify the position where the word ``burger'' appears in the review.}

%\noindent $\bullet$ Emotion \cite{Emotiondataset}: 
% This dataset collects scores for different Emotion from a small piece
% of text, such as anger, disgust, fear, joy, sadness and surprise. 
% For each piece of text, and an emotion (e.g., anger) 
% the worker is asked to give a score in the
% range [0, 100], which indicates the degree of emotion in the text. 
%This is a public dataset based on the affective text annotation task, and each task is answered 10 times.  
\stitle{Emotion~\cite{Emotiondataset}}
This dataset collects scores for different emotions from a small piece of text.
Each worker is asked to give a number in [0,100] for each of the following six emotions: anger, disgust, fear, joy, sadness, and surprise, and a single numeric rating in the interval [-100,100] for his overall (positive or negative) sentiment about the text. Here, all the 7 attributes are continuous. The workers' answers and the ground truth are provided by \cite{Emotiondataset}.
%to denote the overall positive or negative valence of the emotional content of the headline. Thus we have seven attributes(six emotion scores and one overall valence) and they are all continuous. 

%Table \ref{table:dataset_setting} shows the statistics of these three data sets. 

For the Celebrity and Restaurant datasets, we collected the workers' answers by AMT~\cite{AMT}. Each task in Celebrity and Restaurant is answered 5 and 4 times, respectively,  by different workers. We spent $\$0.05$ per HIT where the number of tasks put in a HIT is the same as the number of columns. So, each dataset costed us $\$43.5$ and $\$40.6$, respectively. For Emotion, we use the workers' answers from \cite{Emotiondataset}; each task is answered 10 times. 
%\nikos{unclear if for all tested methods you assign the same tasks to the same workers. specify this.}
%\reynold{Are the 3 real datasets downloadable in public? If so we should include a URL in our references.  Also, can we say that all the source codes for the experiments are available and provide a link?  This will give better impression to reviewers that our experiments are reproducible.} 

\subsection{Truth Inference} 
\label{subsec:TI}

We study the effectiveness of our truth inference approach and other existing solutions:
%previous work with respect to the accuracy of the estimated true values in the three tables, given crowdsourced data from AMT. 
%\subsubsection{Settings} 
%We introduce the competitors to our truth inference approach for record data and the effectiveness measures that we use in the evaluation. 

%\noindent \textbf{Competitors.} 
%\stitle{Competitors}
%We compare with the following methods:\\
\noindent (1) For both categorical and continuous data: \\
\noindent $\bullet$ T-Crowd: our method proposed in Section \ref{sec:TruthInference}.  {\em TC-onlyCate} and {\em TC-onlyCont} are the constrained versions of T-Crowd that apply only on the categorical or continuous attributes. \\
\noindent $\bullet$ CRH \cite{CRH}: CRH detects truth from
heterogeneous data types by minimizing a loss function.\\
\noindent $\bullet$ CATD \cite{CATD}: CATD detects truth
from multi-source data that follows a long-tail distribution along with 
confidence intervals. \\
\noindent (2) For categorical data only: \\
\noindent $\bullet$ Majority Voting (MV): MV determines the correct labels based on the majority of answers from workers. \\
\noindent $\bullet$ D$\&$S \cite{emAlgorithm}: D$\&$S iteratively estimates each worker's confusion matrix, which is used to infer the correct labels. \\
\noindent $\bullet$ GLAD \cite{GLAD}: GLAD is a probabilistic approach for crowdsourcing categorical data.\\
\noindent $\bullet$ Zencrowd \cite{ZenCrowd}: Zencrowd is a variant of D$\&$S. \\
%which infers the correct labels, the quality of workers and the difficulty of tasks simultaneously. \\
\noindent (3) For continuous data only: \\
\noindent $\bullet$ Median: Median uses the median of workers' answers as
the estimated true value. \\
\noindent $\bullet$ GTM \cite{GTM}: GTM is a truth-finding method specially designed for continuous data.

%\noindent \textbf{Effectiveness Measure.}  
\stitle{Effectiveness Measures}
We
adopt the following measures, proposed in \cite{CRH}, for evaluating
the effectiveness of truth inference on categorical and continuous data
items:\\
%Since we have two types of data: continuous and categorical. To evaluate the effectiveness of various truth inference methods, we adopt the following measures for these two data types. They are proposed in \cite{CRH}.\\
\noindent $\bullet$ Error Rate: For categorical data, we measure the Error Rate
by computing the percentage of mismatched values between each method's
predicted truth and the ground truth. \\
\noindent $\bullet$ MNAD (Mean Normalized Absolute Distance): It is
the root of mean squared distance (RMSE) between each method's estimated truth and the ground truth. Since different attributes have different scales, we normalize each attribute's RMSE by its own standard deviation and average them.

\stitle{Effectiveness Comparison}
In Table \ref{table:Effectiveness}, we summarize the effectiveness of truth inference by all methods in terms of Error Rate and MNAD on the three real-world datasets. 
We can observe that our proposed approach T-Crowd is better than all other methods both on categorical data and continuous data. %(it only loses marginally to GLAD on Restaurant Data). 
On Celebrity, our method reduces the error rate by 4\% on categorical data and the MNAD by 2.7\%  on continuous data compared to the best result of other methods. 
The corresponding reductions on Restaurant are 2.6\% and 4\%. 
%On Restaurant, our method reduces the error rate by 2.6\% on categorical data and 4\% MNAD on continuous data.
On Emotion, we outperform previous work by 10\%. 
CRH does not have stable performance as it is effective on Celebrity and Restaurant, but ineffective on Emotion. 
%The performance of CRH is influenced by special dataset. 
Similarly, CATD is good in terms of error rate but not good in terms of MNAD. 
Overall, our method is more robust compared to CRH and CATD.
% \nikos{briefly explain why}.\caihua{Because they are good at some dataset but perform badly in other datasets.}
%  stable compared with them.

We also test constrained versions of T-Crowd that apply only on the categorical or only on the continuous attributes of the table;
%if we only have categorical data \nikos{do you mean applied to only categorical data?}\caihua{only the categorical columns} or continuous data; 
the corresponding versions of our approach are denoted by {\em TC-onlyCate} and {\em TC-onlyCont}, respectively. Note that the effectiveness of T-Crowd 
is better than that of its constrained versions and that the constrained versions are competitive compared to other methods in their class. %(especially {\em TC-onlyCont} has much lower MNAD compared to CRH and CATD).
%slightly superior to the
%is better than that of {\em TC-onlyCate} and {\em TC-onlyCont} except MNAD in Restaurant if we have more data \nikos{what do you mean by `if we have more data'?} \caihua{It means more answers of a worker, so we evaluate his quality more accurately.}

In summary, T-Crowd  outperforms truth inference approaches which are applied on categorical and continuous data separately. This result demonstrates the benefit of 
modeling worker quality by a probabilistic model in a unified manner for all datatypes.

\begin{table}[] 
\centering
\caption{Effectiveness of Truth Inference} \small
\label{table:Effectiveness}
\begin{tabular}{@{}c@{~}c@{~~}c@{~~}c@{~~}c@{~}c@{}}
\toprule
                & \multicolumn{2}{c}{Celebrity} & \multicolumn{2}{c}{Restaurant} & Emotion \\ \cline{2-6}
Method          & Error Rate         & MNAD          & Error Rate         & MNAD           & MNAD         \\
\midrule
T-Crowd     & {\bf 0.0441}       & {\bf 0.6339}  & {\bf 0.1855 }      & {\bf 0.5607}  & {\bf 0.5961}       \\
CRH             & 0.0460             & 0.6737        & 0.1921             & 0.5835         & 0.7224       \\
CATD            & 0.0498             & 0.7113        & 0.1954             & 0.7234         & 0.6648      \\
Maj. Voting     & 0.0573             & /             & 0.2003             & /              & /            \\
EM              & 0.0620             & /             & 0.2463             & /              & /            \\
GLAD            & 0.0498             & /             & 0.1905             & /              & /            \\
Zencrowd        & 0.0479             & /             & 0.1872             & /              & /            \\
TC-onlyCate    & 0.0498    & /             & 0.1986             & /             & /       \\
Median          & /                  & 0.6998        & /                  & 0.6784         & 0.7026       \\
GTM             & /                  & 0.6516        & /                  & 0.5871         & 0.6792       \\
TC-onlyCont    & /         & 0.6400        & /                  & 0.5682        & {\bf 0.5961}   \\
\bottomrule
\end{tabular}
\end{table}

\subsection{Task assignment}
\label{subsec:TA}

\begin{figure*}[]
 \centering
 \includegraphics[width=1 \textwidth]{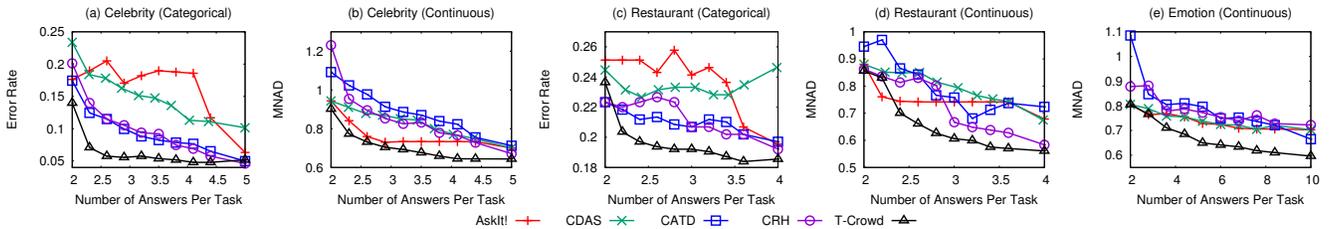}
\caption{End-To-End System Comparison (Effectiveness)}\label{fig:ta} 
\end{figure*}

%\subsubsection{Settings}
We compare the effectiveness of task assignment by our approach against other crowdsourcing methods. 

% We also use the same datasets and we guarantee that each task in $C$ is answered by many times and enough to infer the truth.

% \nikos{the setup of the experiment is unclear. suppose you have $|C|$ tasks and $W$ workers. do you first get the results of all $W$ workers on all $|C|$ tasks and then simulate a task assignment policy by only using the data that are selected by the policy? If this is so, you have to specify this explicitly.} \caihua{I don't have results of all $W$ workers on all $|C|$ tasks because it's too expensive and it's unrealistic(we cannot focus a worker to answer all $|C|$ tasks in actual system.) Thus, we guarantee that each task in $C$ is answered by enough times (5 or 4 times) in our dataset.}

%Because task assignment employs truth inference methods to estimate information gain, we also include CRH and CATD, which are applied on nominal and continuous data, as alternatives for truth inference to our our method method. Since these are not probabilistic approachs, they cannot calculate uncertainty (entropy) or informatiom gain (delta entropy). 

%\noindent \textbf{Competitors.} 
\stitle{Competitors}
We compare T-Crowd, which uses the truth inference method of Section~\ref{subsec:inference} and the task assignment method in Section~\ref{sec:task:corr} with the following approaches: 
%\nikos{I do not understand. What's the point of truth inference if you are not going to use it in estimating information gain. If you are going to select the cells sequentially without any truth inference based decision, what's the point of using CRH or our method? In general, you should present these approaches more clearly. Now they are not understandable.} \caihua{The reason that I use loop+crh and loop+catd is to make up of some baseline. Since crh and catd are suitable for heterogeneous data but cannot use entropy or delta entropy. A Good way is to loop. I add two more baselines which are from \cite{askit} and \cite{cdas} and I change them suitable for my setting.}

\noindent
$\bullet$ CDAS \cite{cdas}: CDAS measures the confidence of the currently estimated values
of all tasks based on a quality-sensitive answering model. Each task for which we are already confident is ``terminated'' and no longer assigned to workers. At each step, CDAS  
selects at random a non-terminated task to assign to the incoming worker. \\
$\bullet$ AskIt! \cite{askit}: AskIt! uses an entropy-like method to define the uncertainty of each task, and infers the truth by Majority Voting. The task with the highest uncertainty is the next one to be assigned to the incoming worker.\\
$\bullet$ CRH \cite{CRH}: CRH is an inference method suitable for heterogeneous data. It does not focus on task assignment, hence, tasks are randomly assigned to the incoming workers.\\
% because it is not probabilistic. \nikos{so, why is this different to a random assignment? is there no way to select which task to assign?} \caihua{Solved.}\\
$\bullet$ CATD \cite{CATD}: CATD is an inference method suitable for heterogeneous data, which does not focus on task assignment. Similar to CRH, we collected answers by randomly assigning tasks. %\nikos{again, this looks  like totally random. so, what's the difference between the two above competitors? They seem to behave the same.} 

%\noindent \textbf{Effectiveness Measure.} 
\stitle{Effectiveness Measures}
As in the evaluation of truth inference, we use Error Rate and MNAD to measure task assignment quality. Specifically, for each tested method, we measure the Error Rate and MNAD as a function of the average number of answers collected by task
so far.
% \nikos{is this average number}.\caihua{Yes} 
A good method would be able to converge fast with fewer answers per task(i.e., by performing fewer assignments and hence spending less money). Besides, it should achieve a lower true value estimation error when it converges.

%\nikos{We need a CLEAR definition of what ``assignment quality'' is. Say that }

%We also test the effectiveness of the HIT size $k$ and the weight of hybrid information gain $\kappa$.

% \begin{figure*}
%  \centering
% 	\subfigure[Celebrity Data]{
%  	\label{fig:ta_cele} %% label for first subfigure
%     \includegraphics[width=0.32\textwidth]{figs/cele.png}}
%   	\subfigure[Restaurant Data]{
%     \label{fig:ta_absa} %% label for second subfigure
%     \includegraphics[width=0.32\textwidth]{figs/absa.png}}
%   	\subfigure[Emotion Data]{
%     \label{fig:ta_emotion} %% label for second subfigure
%     \includegraphics[width=0.32\textwidth]{figs/emotion.png}}
%   \caption{Task Assignment}
%   \label{fig:ta} %% label for entire figure

% \end{figure*}

\stitle{End-To-End Comparison} 
 To perform a fair comparison with existing work, we performed experiments on AMT~\cite{AMT} 
%at the same time 
by using the same settings for the different methods (i.e., each task costs the same). We use the `external-HIT'~\cite{externalHIT} feature provided by AMT to dynamically assign tasks for the incoming worker. To assess the effectiveness of task assignment, we vary the budget and compare the Error Rate and MNAD of each method under the same budget. %we measure the error rate and MNAD of each method.
To be specific, for each budget, we record  the error rate and MNAD on all real datasets as more answers are collected.
%\nikos{maybe you have to mention how many \$\$\$ you spent for the experiment in total. I think they do so in other crowdsourcing papers.}\caihua{mentioned it before.}

% Task Assignment is an iterative process that selects the next task(s) to assign to a worker based on the quality of the existing answers. Since, at first, we do not have any answers, we generate a random initial answer set for all cells of the table (by randomly selecting two answers) and then apply all tested task assignment policies on this initial answer set.
%To test the behavior of different task assignment methods, we start
%from the same initialization answer set. We random select two answers from a cell and make up of initialization answer set, which is the same to all methods.

%Then we select the cell and the worker based on different policies. 
% We compare the quality of the results obtained by the different methods
% after getting the same number of answers from the workers (in other
% words, after using up the same budget).
%In the three real datasets, each cell has only
%two answers at the initial stage and they receive more answers, as the
%budget gradually increases

% \nikos{the budget is not increasing, this is confusing}. \caihua{I mean the number of answer is increasing. Because each answer needs one unit of budget, budget is increasing.}
Figure \ref{fig:ta} shows the experimental results. Naturally, the error rate and MNAD of all assignment policies decrease as more answers are received from the workers and converge to good results after a large number of answers. 
% When comparing policies that pick the next cell sequentially (looping), observe that  
% %When using looping to pick the next cell, our proposed method 
% our method is better than CATD and CRH both on categorical and continuous data in most cases. This is another indication regarding our method's superior effectiveness in truth inference.
% \nikos{unclear to me why this is happening. see my earlier comment.} \caihua{Using looping, CATD, CRH and our method has the same current answer set. But the error rate and MNAD of our method is smaller.}
Askit! uses an entropy-like method, which makes it prefer continuous tasks first. Thus its MNAD drops fast while the error rate remains high. After selecting all continuous tasks, its error rate starts to drop.
Since no task is terminated in the first few iterations, CDAS 
%does not distinguish between terminated and non-terminated tasks at first stage, it 
converges slowly. In addition, since its inference method is simple, the final inferred result is not good compared to that of other methods.
CRH and CATD are not probabilistic, which do not use metrics, like entropy or 
%a probabilistic model in the assignment, so they do not calculate 
information gain, as the objective for task assignment, so they do not perform as well as T-Crowd. They are superior to Askit! and CDAS because they are more effective in inferring the true values of tasks.

We observe that our proposed approach T-Crowd converges much faster to a low error rate and MNAD compared to the other policies. Specifically, T-Crowd
%The error rate and MNAD drop fast and 
converges to low values before 
the average number of answers per task is 3 on Celebrity and Restaurant and 6 on Emotion, which shows the effectiveness of our structure-aware information gain measure as an assignment criterion. 
In addition, due to  our superior truth inference method, the values eventually inferred by our framework are better compared to those inferred by the other methods.

%By using ra, they do not perform well in task assignment. 

%\nikos{why is there a difference between these two methods if they both use exactly the same looping policy for selecting the next task?} \caihua{They both use loop so their answer sets are the same. But inferred method is different, so error rate and MNAD is different.}

%\nikos{There seems to be a gap because StructCrowd before (i.e. section 4.1) and after using correlation (i.e., section 4.2) is not compared. If the result is not bad, you should add it.} \caihua{I compare the assignment methods in the section 4.1 and 4.2 in the following(Comparison with Assignment Heuristics.). I just use one dataset because the performance on other two datasets is not significant. Maybe we can change the order of comparison of only assignment methods and comparison of end-to-end methods}
%\nikos{fix the remaining section.}\caihua{fixed}

\subsection{Case Studies}
\label{subsec:case}

We performed several case studies in order to assess the quality of our system. Due to space constraints, we only report the results on Restaurant. Similar observations can be derived by experimentation on the other datasets.

\subsubsection{Worker Quality} Our first study's goal 
%Our goal 
is to show that (1) each worker's actual quality (computed based on the ground truth) is consistent among different attributes; (2) each worker's estimated quality can be well calibrated to the worker's actual quality. 

\begin{figure} 
 \centering
 \includegraphics[width=0.45 \textwidth,bb=0 0 8082 2610]{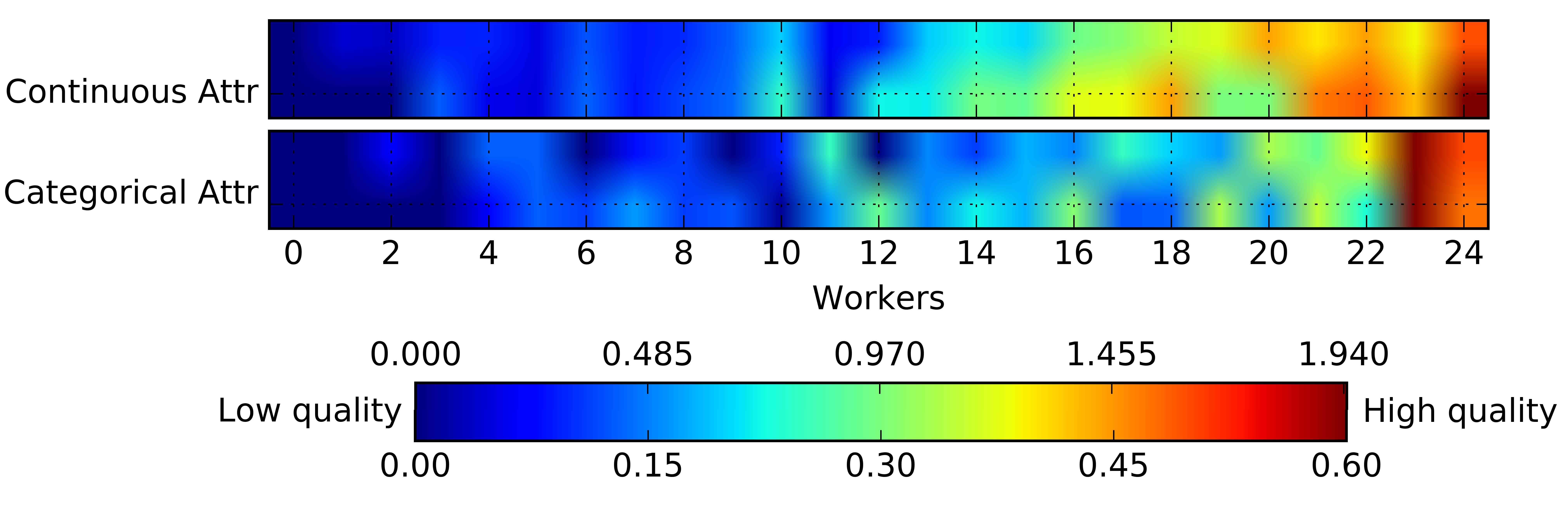}
\caption{Uniform Worker Quality}\label{fig:workerstd}
\end{figure}

\begin{figure}
 \centering
\subfigure[categorical]{
    \includegraphics[width=0.2\textwidth]{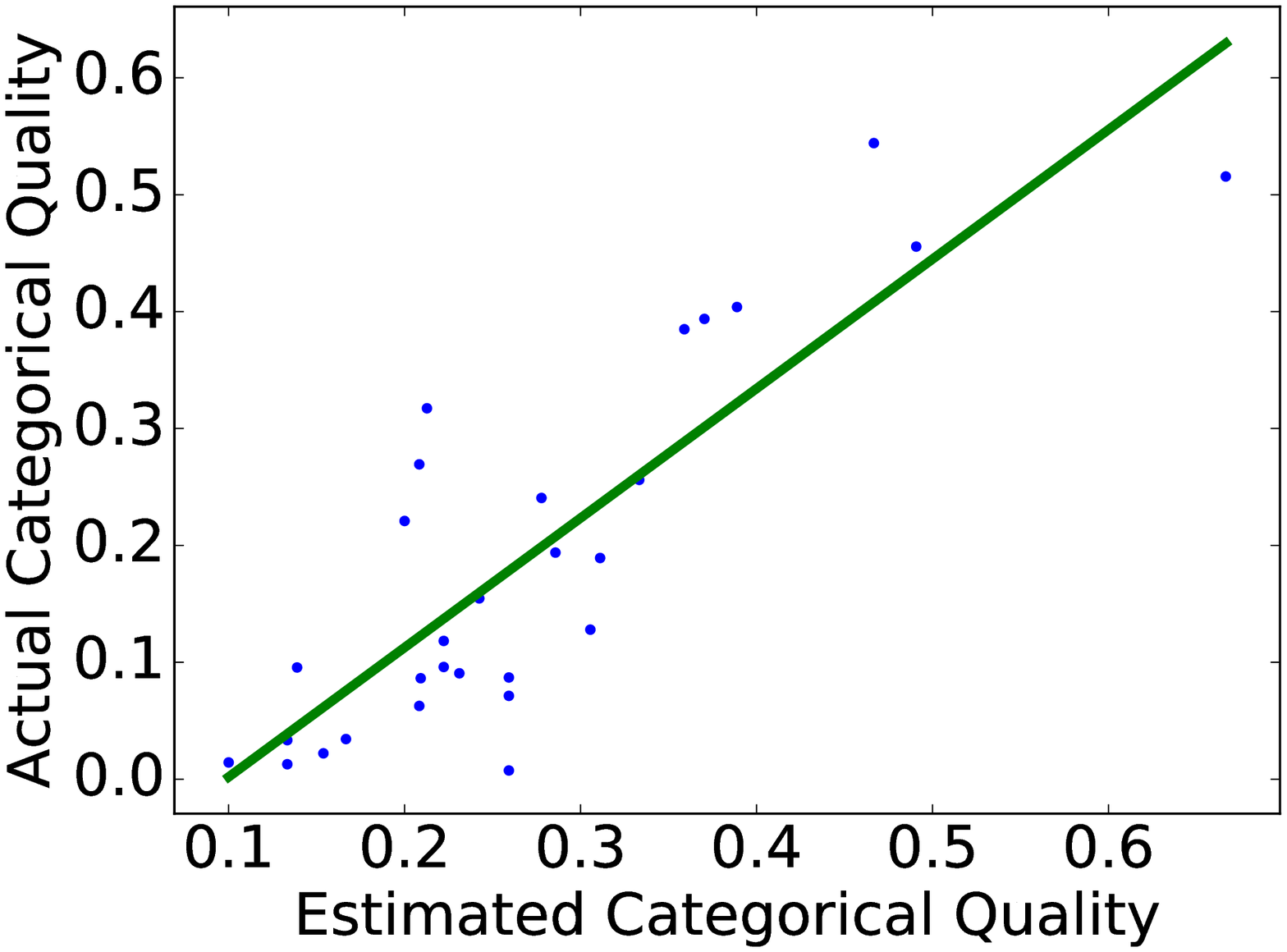}}
  \subfigure[continuous]{
    \includegraphics[width=0.2\textwidth]{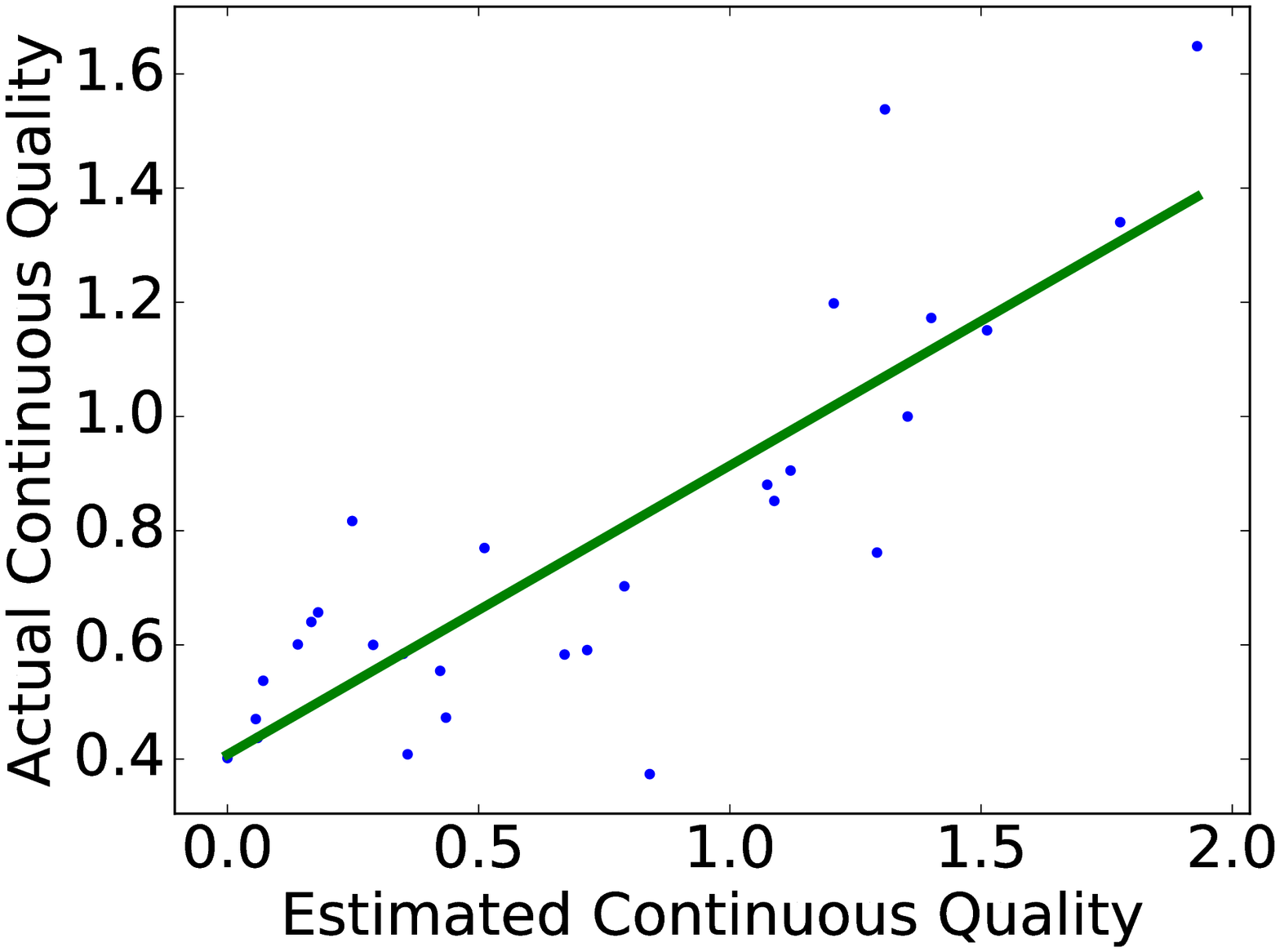}}
  \caption{Estimated and Actual Worker Quality}
  \label{fig:estimatequality}
\end{figure}

\stitle{Consistent Quality for Different Attributes}
We collected statistics from the Restaurant dataset to support our assumption in truth inference: a worker has consistent quality over different datatypes of attributes.
In Figure \ref{fig:workerstd}, we plot a heat map, with the x-axis 
representing 
the 25 workers 
%$u \in U$ 
who have given the largest number of answers 
and the y-axis representing 
categorical attributes `Aspect' and `Sentiment' 
 and continuous attributes `StartTarget' and `EndTarget'.
% for 
%attributes $j \in M$. 
Different colors are aligned to standard deviation values (above the colorbar) for continuous attributes and error rates (below the colorbar) for categorical attributes. 
%is colorbar, with the values above it denoting the standard deviation for continuous attributes while the value below it denoting the error rate for categorical attributes. 
%Values 0 to 24 on the x-axis correspond to 25 workers who have given the largest number of answers in this dataset (each of these workers have at least 5 answers). The values on the y-axis correspond to different attributes, and we show `Aspect' and `Sentiment' for categorical attributes and `StartTarget' and `EndTarget' for continuous attributes. 
The color of each pixel represents the average error of answers given by worker $u$ to the tasks on attribute $j$. For a categorical attribute $j$, the error is the percentage of wrong answers. For a continuous attribute $j$, 
%since we model worker quality as variance for continuous data, 
the error is the standard deviation of the differences between the answers and the ground truth.
%errors of answers given by worker $u$ on attribute $j$. 
%We compute the quality based on comparing each worker's answers with the ground truth. 
The red color (far right) implies larger error and lower worker quality, while the blue color (far left) means smaller error and better worker quality. Observe that the workers have consistent performance for categorical and continuous attributes. In addition, the colors for the same worker are similar regardless the attribute type,
%on the same types of attributes or different types of attributes is very close, 
which means that each worker's actual quality is consistent among different attributes. 

%\noindent \textbf{Calibration to the Actual Quality.} 
\stitle{Calibration to the Actual Quality}
Figure \ref{fig:estimatequality} shows that our estimated quality of a worker is close to the actual quality. Each point represents a worker and the x-axis value is the quality estimated by our method while the y-axis value is the actual worker's quality. We also show the result of a linear regression. Observe the strong correlation between 
%We can see that the correlation coefficients between 
our estimated quality and actual quality; the correlation coefficient is 0.844 for categorical attributes and 0.841 for continuous attributes. 
%Thus, our method is well calibrated to the actual quality of the workers.

%\subsubsection{Case Study on Dataset Restaurant.} Next we perform case studies to observe (1) the comparisons of different assignment heuristics; (2) the correlation of different attributes. Note that due to space limits, we only show the experiments on Restaurant. The same observations can also be found in other datasets.

\begin{figure} [!ht]
 \centering
 \includegraphics[width=0.45 \textwidth]{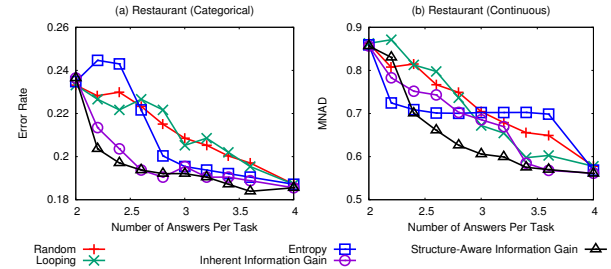}
\caption{Effectiveness of Assignment Heuristics}\label{fig:selfassignment}
\end{figure}

%\vfill\null %for column break
\subsubsection{Assignment Heuristics}
%\noindent \textbf{Comparison of Different Assignment Heuristics.} 
We evaluate the performance of different assignment heuristics. Note that for all of them, we use our inference approach (Section~\ref{subsec:inference}). The tested heuristics are listed as follows: \\
$\bullet$ Random: randomly chooses the task assigned to the worker. \\
$\bullet$ Looping: selects the next task in a round-robin manner. \\
$\bullet$ Entropy: greedily chooses the next task which has the highest uncertainty (defined as entropy).\\
$\bullet$ Inherent Information Gain: proposed in Section~\ref{sec:inherentIG}. \\
$\bullet$ Structure-Aware Information Gain: proposed in Section~\ref{sec:task:corr}. 
%$\bullet$ Hybrid Informatiom Gain: a linear combination of Inherent and Correlation-aware Informatiom Gain (i.e. Eq. \ref{eq:linear} at the end of Section~\ref{sec:task:corr}).

Figure \ref{fig:selfassignment} presents the Error Rate and MNAD as a function of number of tasks assigned to the workers on Restaurant.
The results on the other datasets are similar and omitted for the interest of space.
%due to space constraints.
% w.r.t. different number of answers per task. 
Random and Looping select tasks without considering the answers collected so far, so they converge slowly. Entropy is biased toward selecting continuous tasks over categorical first; hence, this heuristic reduces the MNAD fast, but not the Error Rate. Inherent and Structure-Aware Information Gain consider the continuous and categorical tasks fairly and decrease the Error Rate and MNAD simultaneously. Besides, Structure-Aware Information Gain converges faster than Inherent Information Gain w.r.t. MNAD because it also considers the correlations between attributes. Recall that we use Structure-Aware Information Gain as our default method.

\begin{figure}
\begin{minipage}[c]{.5\linewidth}
\centering
\label{table:corr_sentiment_aspect}
\scalebox{0.7}{
\begin{tabular}{|c|c|c|}
\hline
\diaghead{\theadfont Aspect   Sentiment}%
{Aspect}{Sentiment}&
\thead{correct}&\thead{wrong}\\    \hline
correct & 589   & 90   \\ \hline
wrong & 98    & 35    \\ \hline
\end{tabular}
}
\end{minipage}
%\hfill
\begin{minipage}[c]{.4\linewidth}
\centering
\label{fig:corr_bimul}
\includegraphics[height=80pt]{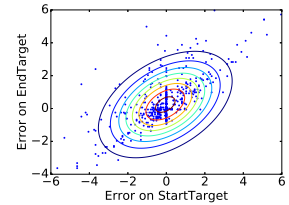}
\end{minipage}%
\caption{Correlation Among Attributes}
\label{fig:correlation}
\end{figure}

\subsubsection{Correlation Among Attributes} 
We perform one more experiment to support 
our assumption that there exist correlations among attributes, 
by analyzing the answers of workers. 

Figure \ref{fig:correlation} shows the experimental results.
%We now test the correlation among attributes on Celebrity. 
In the left part of the figure observe 
%Figure \ref{fig:correlation}, we show 
that attributes `Aspect' and `Sentiment' have strong correlation. Specifically, if a worker answers `Aspect' correctly, the probability to answer `Sentiment' correctly is 86\%. However, if a worker answers `Aspect' wrongly, the probability to answer `Sentiment' correctly is only 73\%. 
In the right part of the figure, we plot a scatter diagram, with each point representing a worker's error on attributes `StartTarget' and `EndTarget'. 
We use {\em maximum likelihood estimation} to obtain the joint distribution of errors on these two attributes as described in Section \ref{sec:task:corr}. We observe a positive correlation between errors on attributes `StartTarget' and `EndTarget', which justifies our proposed Structure-Aware Information Gain method that considers correlations among attributes. For example, if the error of `StartTarget' is 0, the distribution of `EndTarget' error is $\mathcal{N}(0.28,0.76)$. However, if the error of `StartTarget' is 6, the distribution of `EndTarget' error is $\mathcal{N}(3.75, 0.76)$. In other words, knowing the exact answer of a worker on one attribute can help to predict his answer distribution for other attributes better.

% If we know any one answer of them, the distribution of other one answer will change larger than original one.
%if a worker's answer of `name' is correct, his error for `height' has a much smaller variance compared to the case where the answer of `name' is wrong.
%is closer to 0. And if a worker's answer of `name' is wrong, his error of `height' is uniform from -2 to 2.  

% \begin{table}[]
% \centering
% \caption{Correlation between name and nationality}
% \label{table:corr_name_nationality}
% \begin{tabular}{|c|c|c|}
% \hline
% \diaghead{\theadfont name   nationality}%
% {name}{nationality}&
% \thead{correct}&\thead{wrong}\\    \hline
% correct & 615   & 110   \\ \hline
% wrong & 66    & 79    \\ \hline
% \end{tabular}
% \end{table}

% \begin{figure}
%  \centering
% \subfigure[answer of name is right]{
%  \label{fig:subfig:a} %% label for first subfigure
%     \includegraphics[width=0.23\textwidth]{figs/name_height_right.png}}
%   \subfigure[answer of name is wrong]{
%     \label{fig:subfig:b} %% label for second subfigure
%     \includegraphics[width=0.23\textwidth]{figs/name_height_wrong.png}}
%   \caption{Correlation between name and height}
%   \label{fig:tcorr_name_height} %% label for entire figure
% \end{figure}

\subsection{Synthetic Data}\label{sec:exp:synth}
%We first describe the experimental settings and measures, then evaluate the correctness of our method, and finally examine the performance the online assignment algorithms. %\subsubsection{Settings}
In this section, we use two types of synthetic data, in order to test the performance of
our truth inference approach in cases not covered by the real data settings.

\subsubsection{Tests on tables with different properties}
We assess the performance of T-Crowd
in terms of truth inference effectiveness 
by changing the following 
parameters of our data generator:
%for data of different characteristics.
% various situations of datasets. 
%Our synthetic data generator takes three parameters: 
the number of columns $M$, the ratio of categorical 
to the total number columns $R$ and the average difficulty of tasks $\mu\{\alpha_i\beta_j\}$.  
The default parameters are $M = 10$, $R = 0.5$ and $\mu\{\alpha_i\beta_j\} = 1$.
%\nikos{you've used $D$ to mean something else before. better use another symbol. }
The rest of the settings are as follows:
%We want to explore how they influence the performance of our method. 
%\paragraph{Settings}

\textbf{Worker Sequence and Worker Quality:} 
We use the same number of workers as that in our
real experiments for the 
dataset Celebrity and assume that the workers arrive in the same sequence and that they have the same quality as in the real experiment.
%This is the order by which the workers
%come and their quality. We use a real-world worker sequence obtained by from the real 
%dataset `Celebrity' we used before. 
%We also set  each workers' actual quality in real 
%dataset `Celebrity'. Because we use worker sequence on `Celebrity', the number of workers is fixed.

\textbf{Data and Ground Truth Generation:} We implemented a generator for a table that takes as input the number of rows $N$
and columns $M$, and the datatype and domain range of each column.
The number of possible answers in a categorical column is generated from a uniform distribution $U(2,10)$. The domain of a continuous column is $[0, 1000]$.
The ground truth $T_{ij}$ of each cell $c_{ij}$ is generated by
selecting a value in the corresponding domain randomly.

%\nikos{uniformly, or by a normal distribution for a continuous  attribute? do you have differentprobabilities for the values of categorical attributes?}. \\ 

\textbf{Workers' Answers:} For each worker in sequence, his answer at each cell needs to be generated. The answer $a_{ij}^{u}$ of each worker $u$ at each cell $c_{ij}$ is
created based on the ground truth $T_{ij}$ and his quality $q_{u}$, based on Eq. \ref{eq:contap} and \ref{eq:cateap}.

%\textbf{Assignment Strategy:} 
%Assignment strategy controls selecting which set of cells when a worker comes. 
For fairness to all methods and since we focus on truth inference, we simulate the assignment strategy used in AMT, i.e., each task gets the same number of answers.
%, which is fair to each method.
%We set the number of cell in a batch $k$ as the number of columns. If we have more than $k$ cells which have the same minimum number of answers, we random select $k$ cells. 
%\nikos{unclear. say if all compared methods use exactly the same assignment strategy.} \caihua{Solved.}
For different parameters, we generate new datasets one hundred times and average the results to obtain the error rate and MNAD. We also run other inference methods and found that our method is dominant both on error rate and MNAD. %Thus, we only show our method's result. \nikos{but I thought that the objective is to compare our inference method to other methods like CRH here. So this is confusing.}

\begin{figure}[!ht]
 \centering
\subfigure[Categorical Columns]{
    \includegraphics[width=0.22\textwidth]{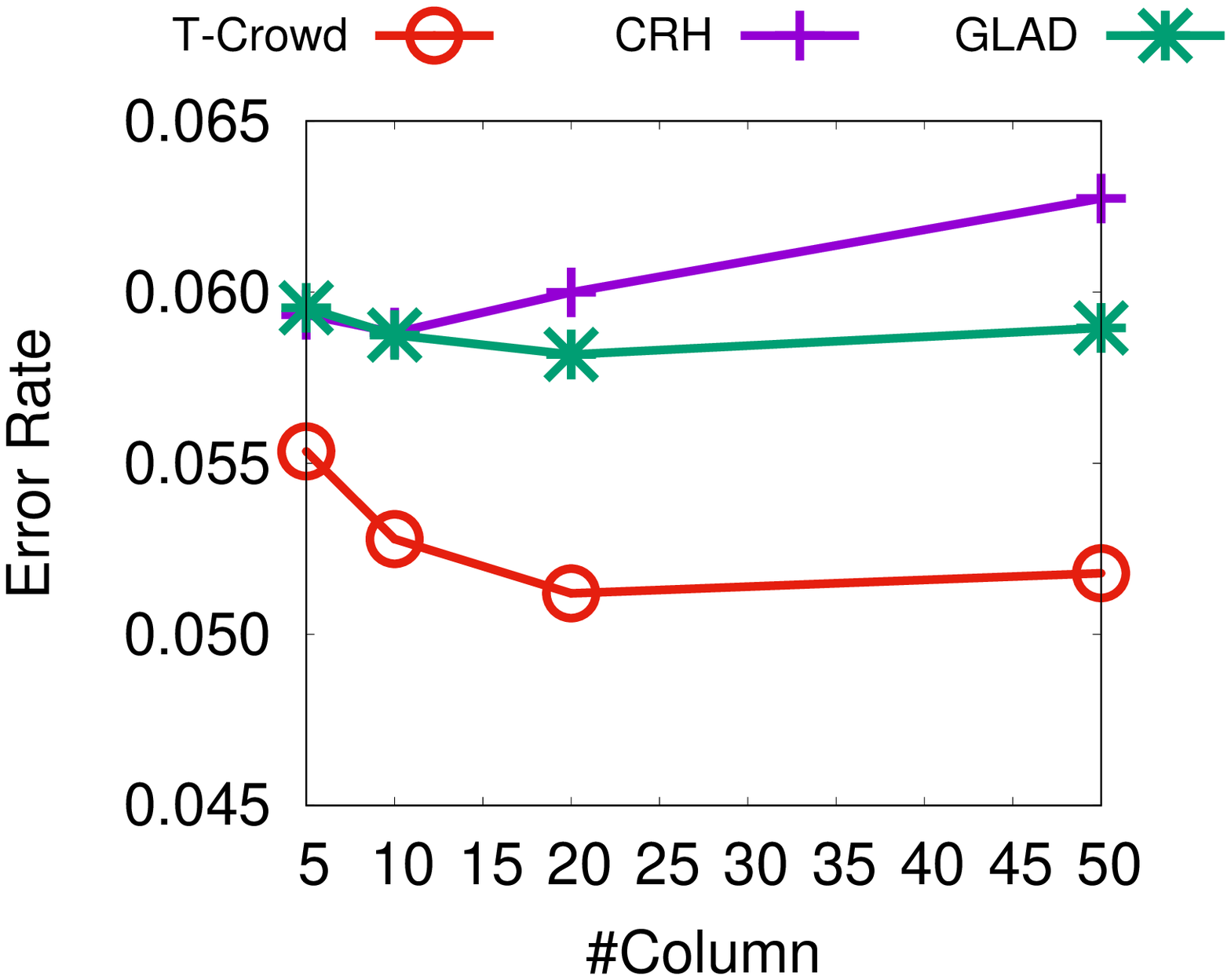}}
  \subfigure[Continuous Columns]{
    \includegraphics[width=0.22\textwidth]{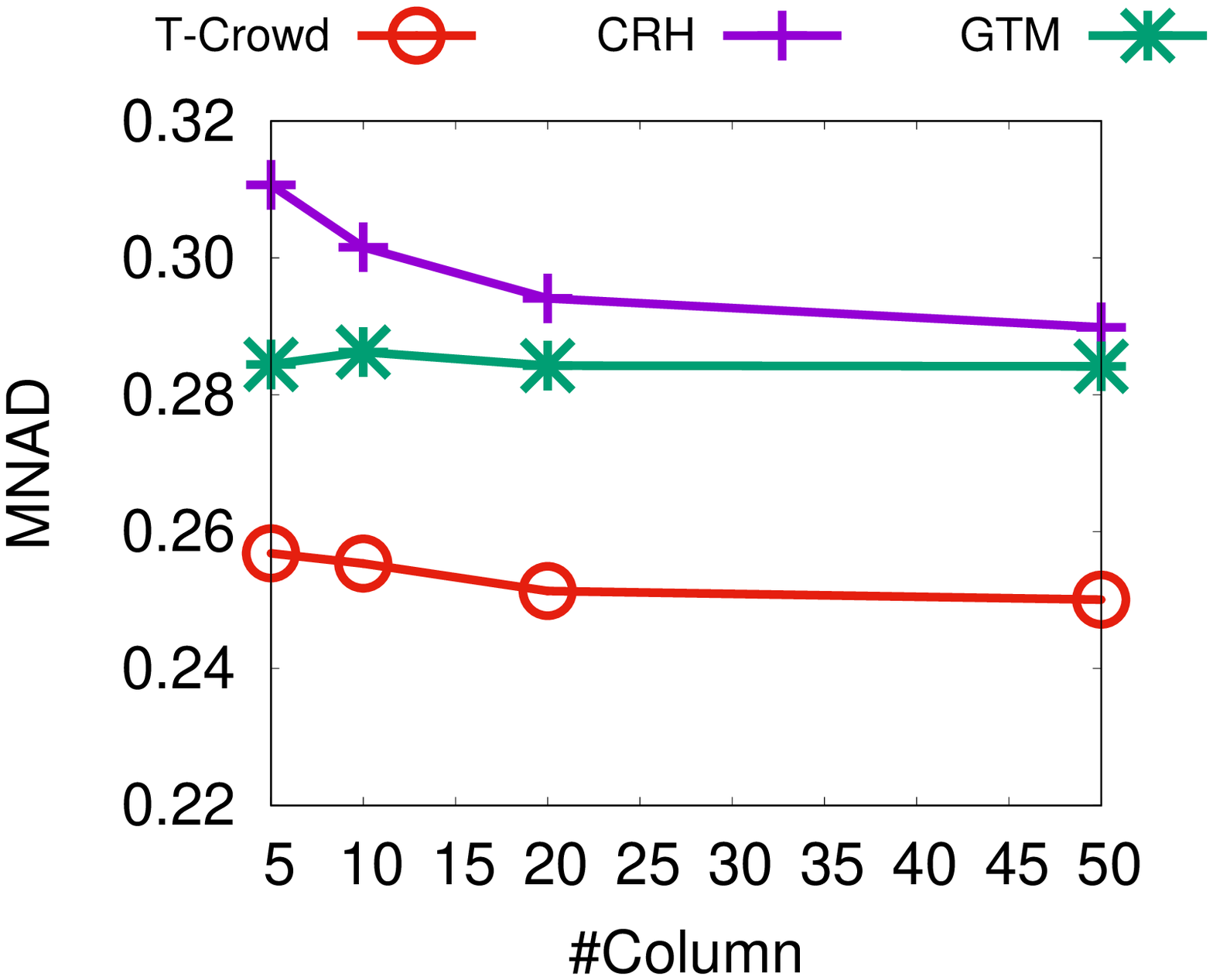}}
  \caption{Effect of the Number of Columns}
  \label{fig:simulated_col_vary} %% label for entire figure
\end{figure}

\begin{figure}[!ht]
 \centering
\subfigure[Categorical Columns]{
	\label{fig:simulated_ratio_error_rate}
    \includegraphics[width=0.22\textwidth]{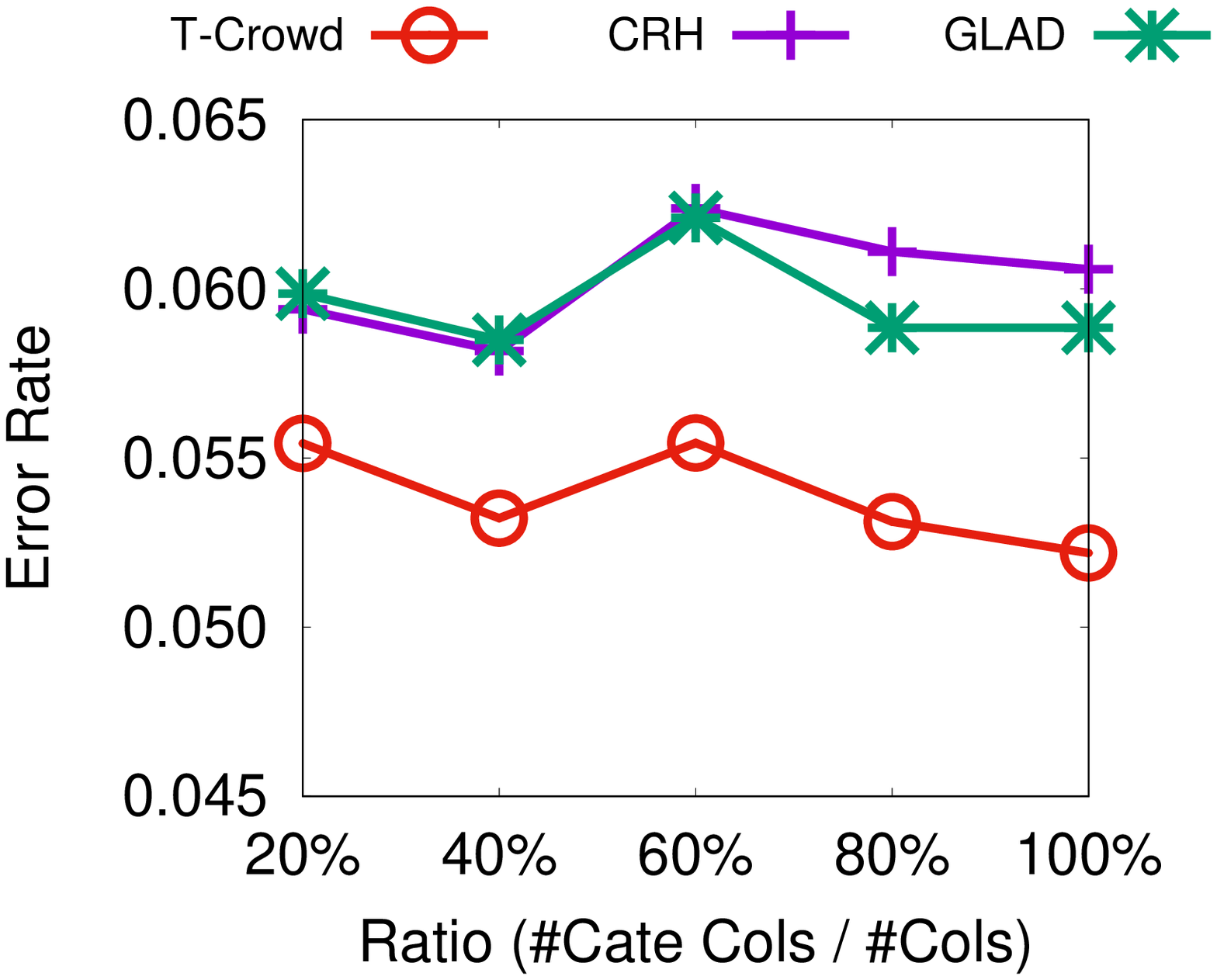}}
  \subfigure[Continuous Columns]{
  	\label{fig:simulated_ratio_mnad}
    \includegraphics[width=0.22\textwidth]{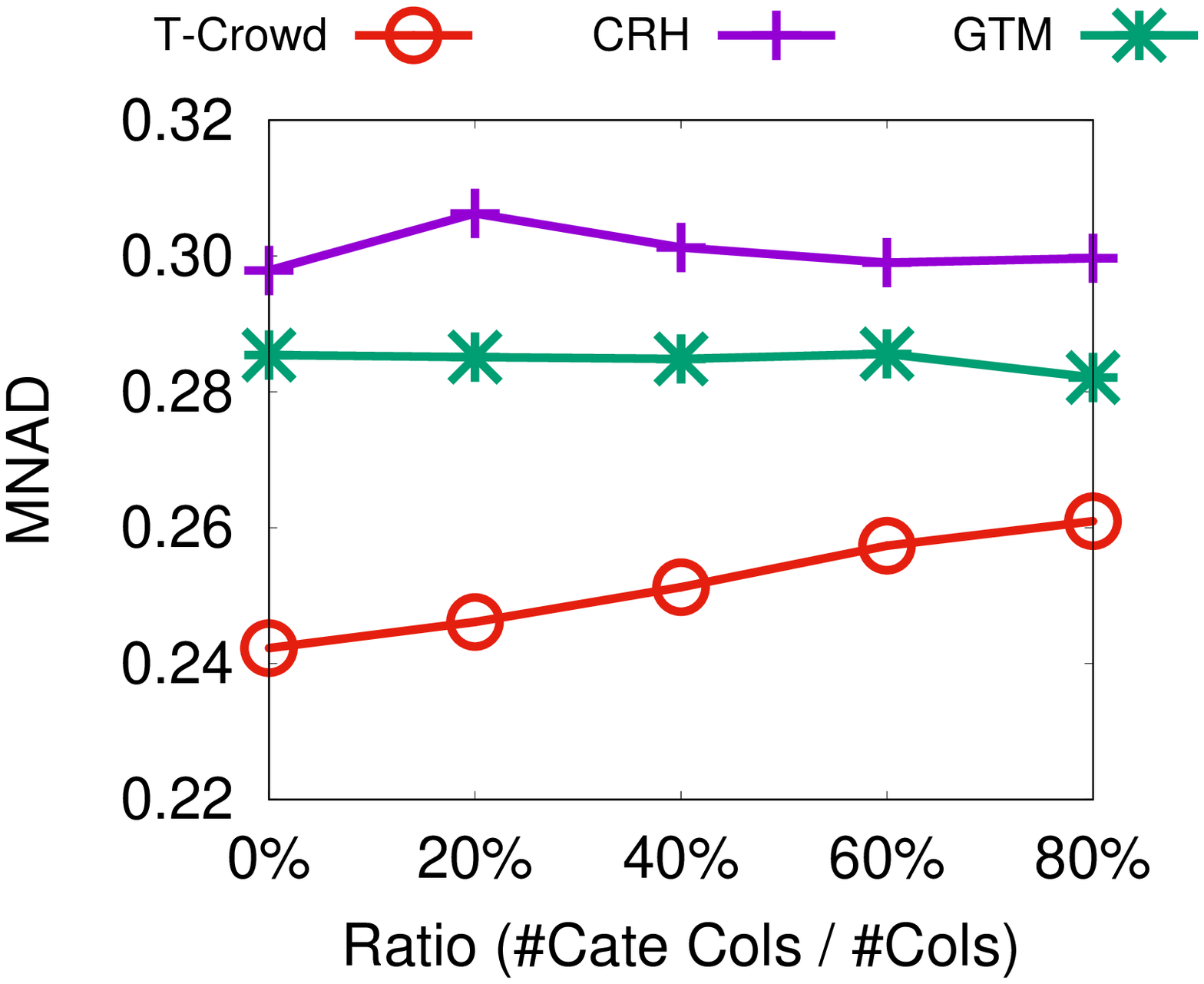}}
  \caption{Effect of Ratio of Columns}
  \label{fig:simulated_ratio_vary} %% label for entire figure
\end{figure}

\begin{figure}[!ht]
 \centering
\subfigure[Categorical Columns]{
    \includegraphics[width=0.22\textwidth]{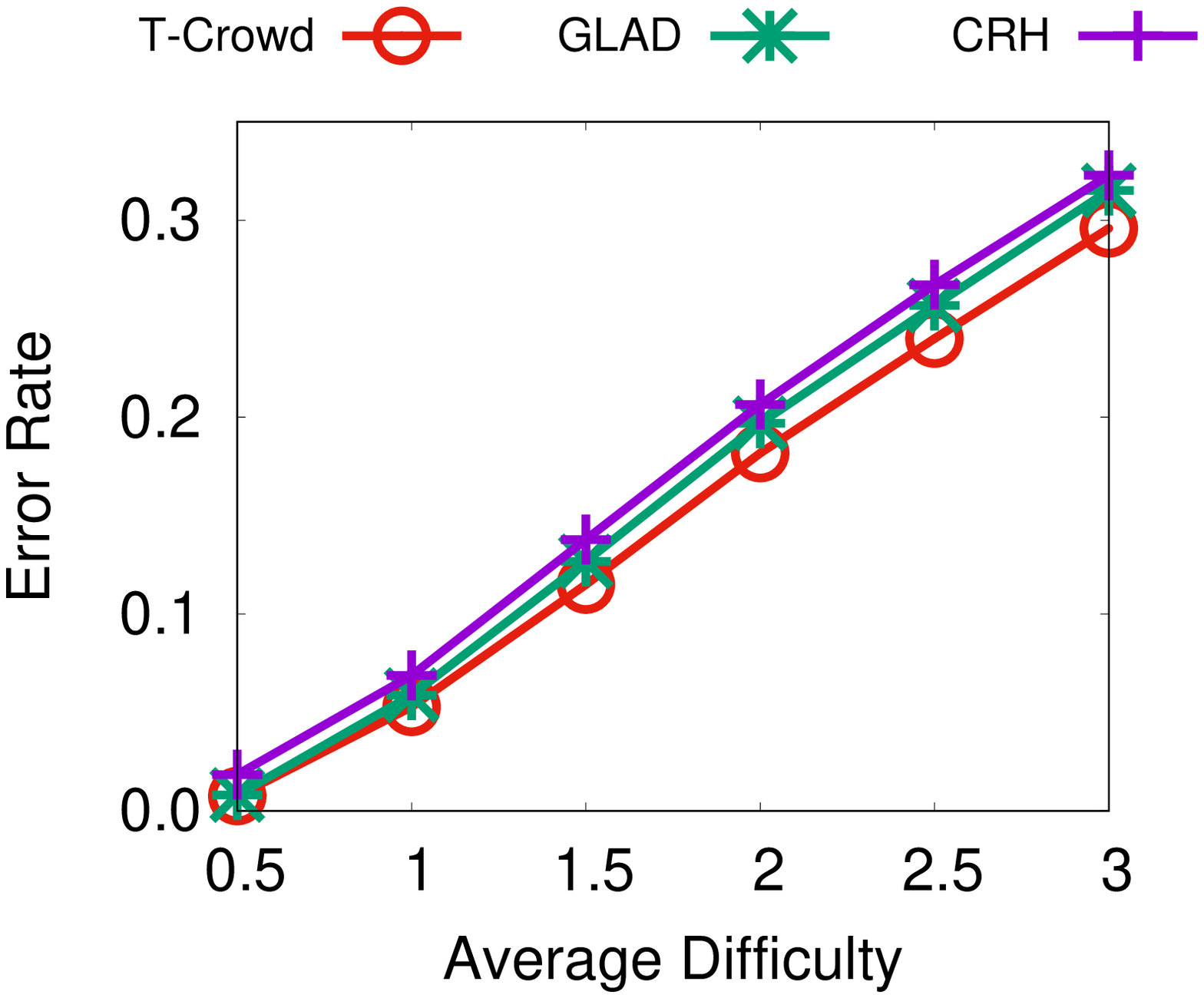}}
  \subfigure[Continuous Columns]{
    \includegraphics[width=0.22\textwidth]{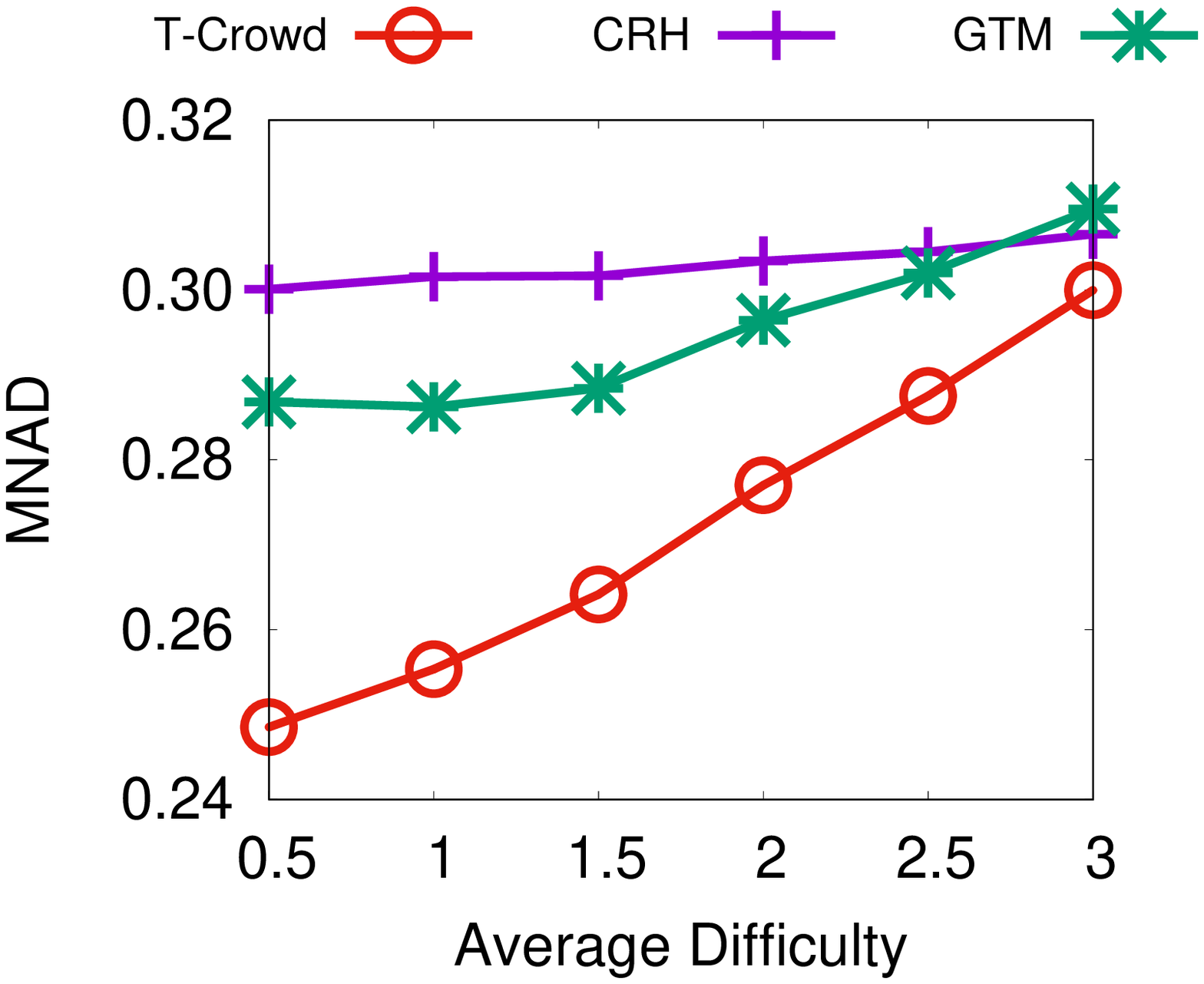}}
  \caption{Effect of Average Difficulty  }
  \label{fig:simulated_ratio_dif} %% label for entire figure
\end{figure}

\begin{figure*}[!ht]
  %\centering  
  \hspace{-20pt}
  \begin{minipage}[b]{0.8\textwidth} 
    \centering 
\subfigure[Error Rate]{
	\label{fig:noisy_error_rate} %% label for first subfigure
    \includegraphics[height=80pt]{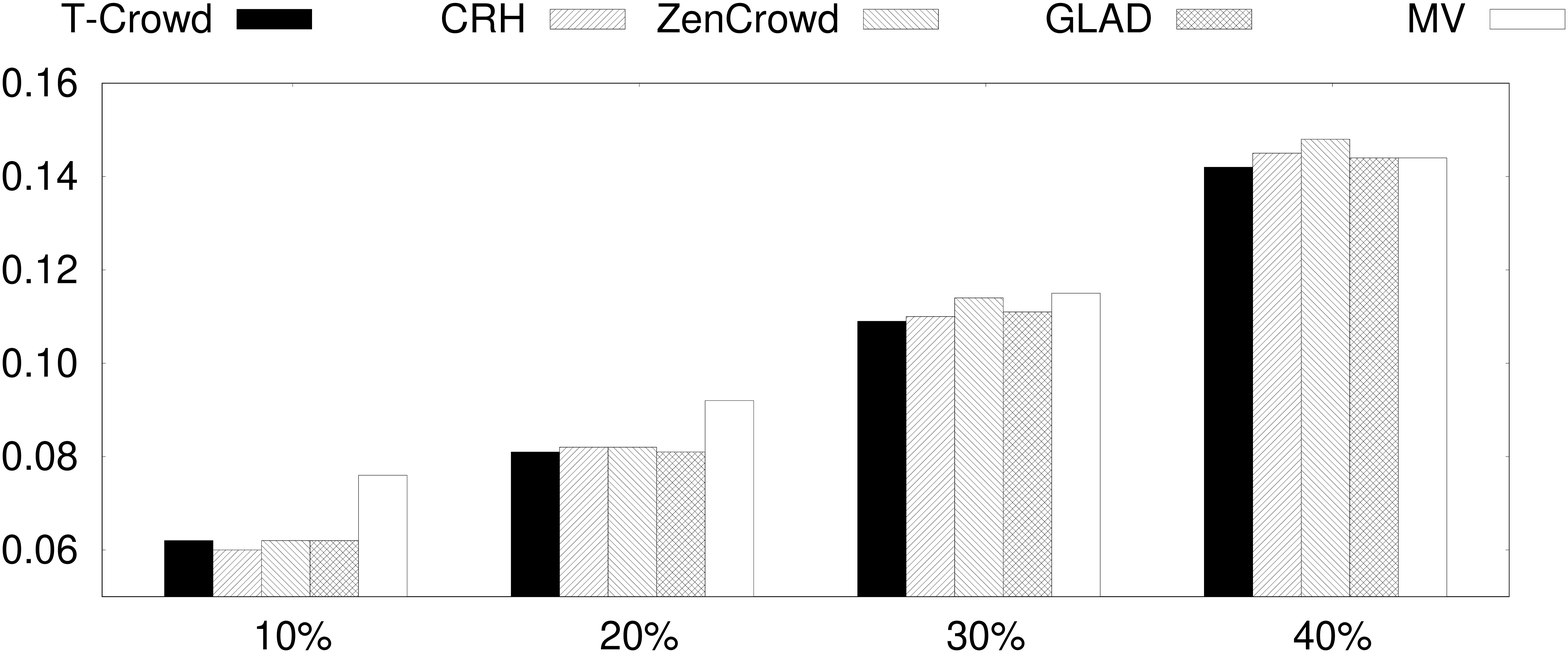}}
  \subfigure[MNAD]{
   \label{fig:noisy_mnad} %% label for first subfigure
    \includegraphics[height=80pt]{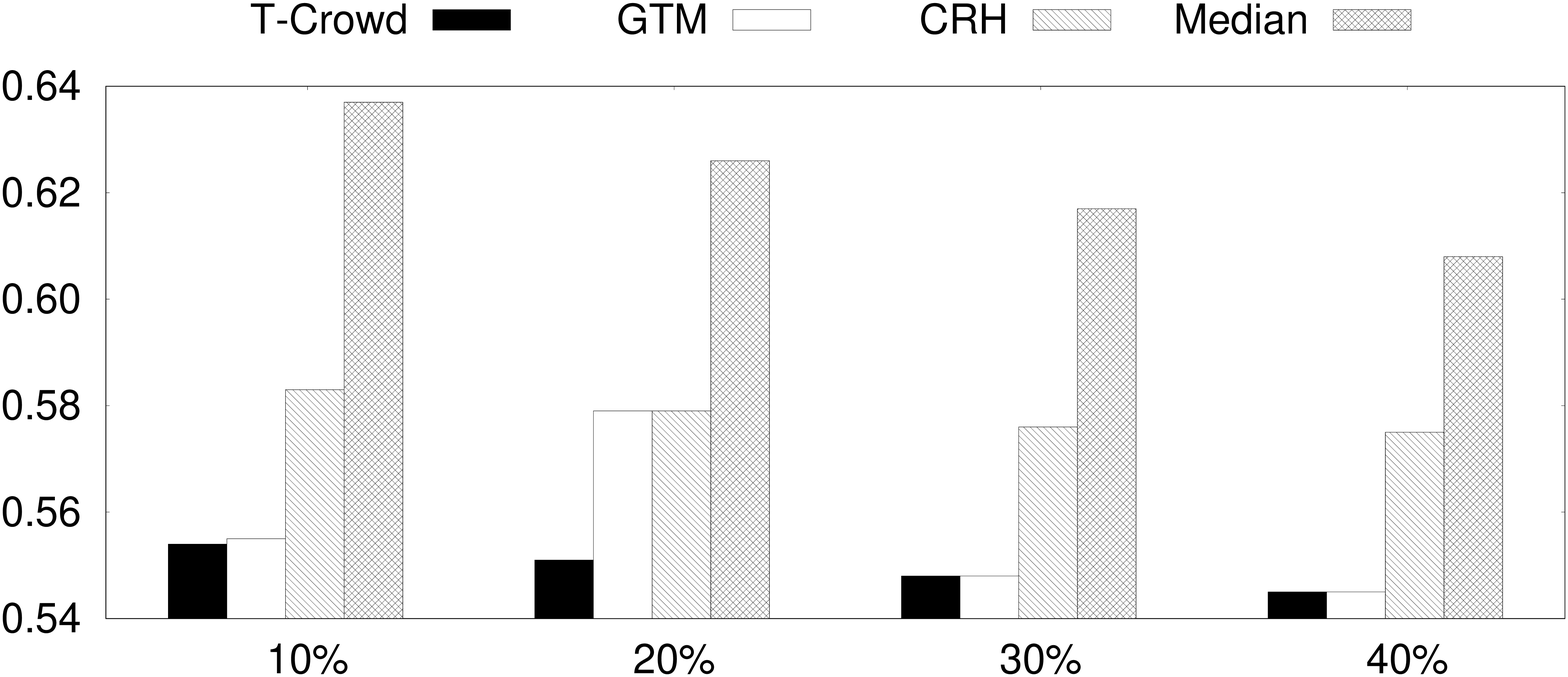}}
  \end{minipage} 
  \begin{minipage}[b]{0.17\textwidth} 
  %  \centering 
    \includegraphics[height=95pt]{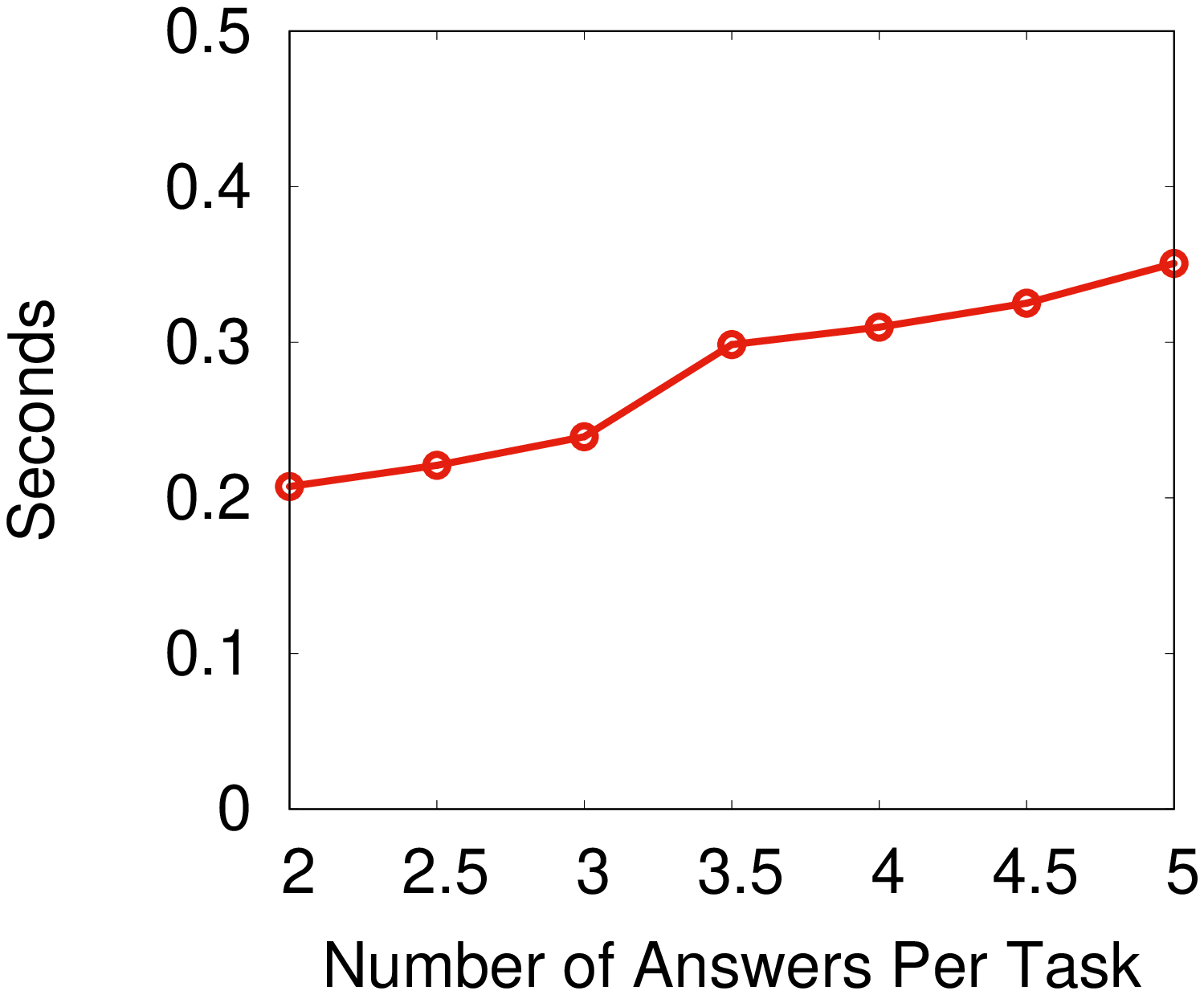} 
  \end{minipage}
  \\[2pt] 
  \begin{minipage}[t]{.69\textwidth}
    \caption{Noisy Datasets} 
  \end{minipage}
  \hspace{-20pt}
  \begin{minipage}[t]{.33\textwidth} 
  	  \caption{Efficiency of Assignment} \label{fig:eff_ass}
  \end{minipage}% 
\end{figure*}

\stitle{Results}
%\textbf{The number of columns $M$:} 
In the first experiment, we vary the number of columns from 5 to 50. Figure \ref{fig:simulated_col_vary} shows that the error rate and MNAD decline gradually when the number of columns increases, showing that T-Crowd infers the quality of each worker and estimates truth more accurate if we have more data. Besides our method is significantly better than the other two approaches.
%\textbf{The ratio of categorical and continuous columns $R$:} 
Next, we vary the ratio of categorical attributes from 0\% to 100\%.
%Thus there is no point in 
Figures \ref{fig:simulated_ratio_error_rate} 
and
%because it does not have categorical data. Similarly, 100\% means that all the tasks are categorical so there is no point in 
Figure \ref{fig:simulated_ratio_mnad}
%. These two figures 
show that our method's error rate and MNAD do not change much when the ratio varies.
%\textbf{The average difficulty of tasks $D$:} We 
Finally, we vary the 
average difficulty of each cell $c_{ij}$ (i.e., the average $\alpha_i\beta_j$, as defined in Section \ref{sec:difficulty of a cell})  from 0.5 to 3. High difficulty implies that the probability that workers answer correctly decreases,
hence the error rate and MNAD increase as shown in 
 Figure \ref{fig:simulated_ratio_dif}.
% shows that error rate and MNAD increases when average difficulty increases. 
For easier tasks, our method is 
significantly better than the others, but when the average difficulty is high, which means that the workers' answers are not credible, all methods perform badly.  
%\nikos{but in categorical attributes the difficulty does not affect the difference between the methods.} \caihua{The reason is range of y-axis. The gap of error rate of different method is 0.005....}

\subsubsection{Noise in Workers' Answers}
To further demonstrate the advantage of our proposed approach T-Crowd, we conduct simulation experiments by adding noise to the original data collected for Celebrity dataset. 
%We use the original settings, including the schema of table, the ground truth, and the same workers. The only difference is that we add different levels of noise into the workers' answers and generate new noisy answers. 
We vary the percentage $\gamma$ of altered original answers by the workers from 
%set the level of noise $\gamma$ from 
10\% to 40\% (i.e., $\gamma$ is the percentage of answers with added noise). 

For a categorical answer, we randomly select a new label from its domain and replace the original label. For a continuous answer, Gaussian noise is added. We first preprocess this answer by transforming it into its z-score. A new normalized answer is generated by adding the noise which was generated by a Gaussian distribution $\mathcal{N}(0, 1)$. We finally change it to the original scale and obtain the new answer. We randomly choose $NM\gamma$ answers with replacement to add noise and the rest of the answers stay the same. 

For different levels of noise $\gamma$, we generate new datasets one hundred times. For each method, we run experiments three times to smoothen out possible instabilities.
%avoid that some result of methods is not stable. 
Hence we run in total 300 simulations for each method and average them to obtain the error rate and MNAD for  different levels of noise $\gamma$. 

Figure \ref{fig:noisy_error_rate} and \ref{fig:noisy_mnad} show the results. The error rate increases while MNAD declines when $\gamma$ increases. The reason for the decrease of MNAD is that the normalization denominator is the standard deviation of answers in each column. The growth rate of standard deviation is higher than that of RMSE which makes MNAD to decline.

T-Crowd performs well and stably when the level of noise $\gamma$ increases both in terms of error rate and MNAD. T-Crowd has a very similar error rate and MNAD to CRH and GTM, respectively.
% on MNAD and better than other methods except error rate when $\gamma = 1$. 

\subsection{Efficiency}
\vspace{-5pt}
\begin{figure}[!ht]
 \centering
\subfigure[Convergence Rate]{
	\label{fig:convergence} %% label for first subfigure
    \includegraphics[width=0.22\textwidth]{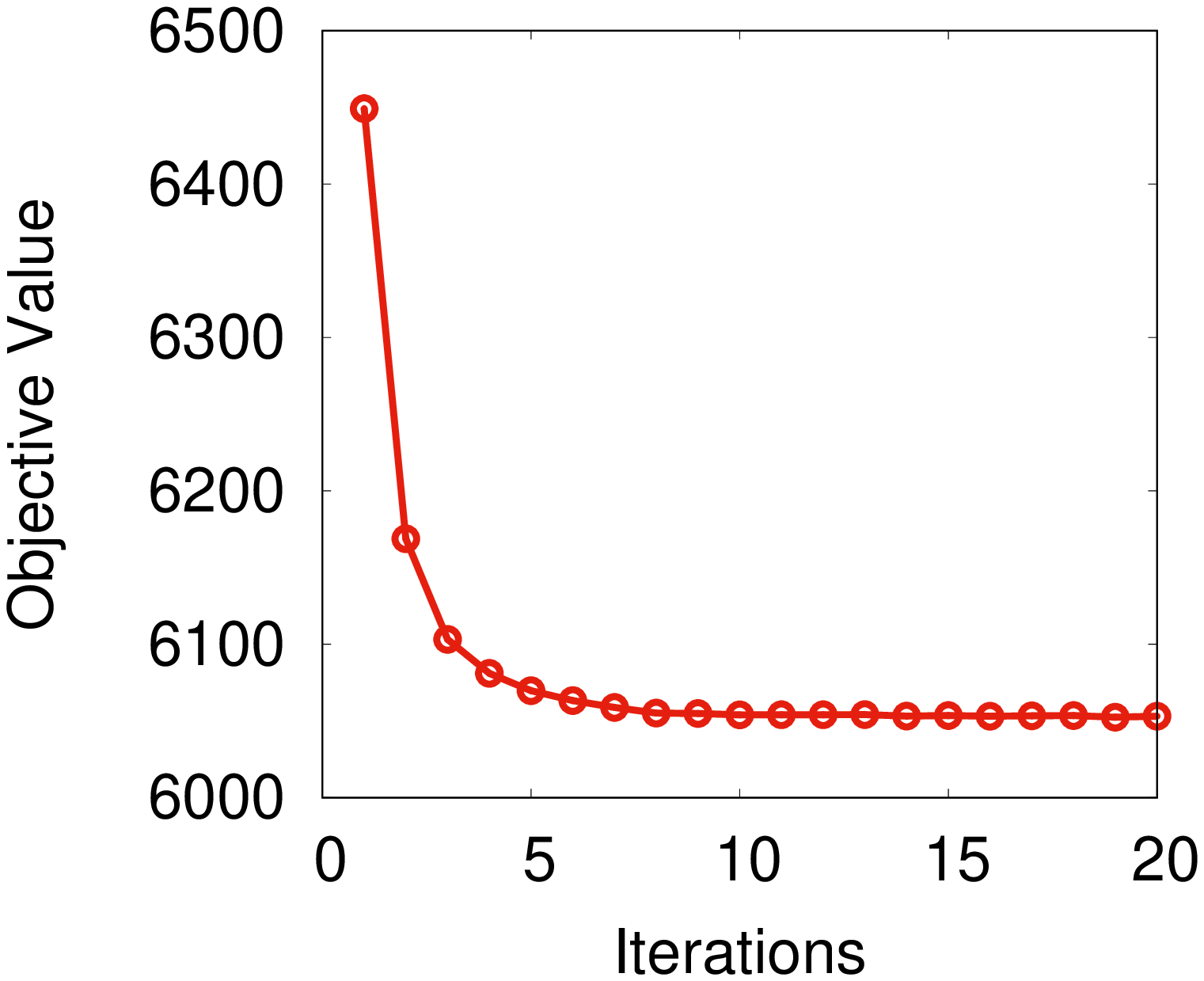}}
  \subfigure[Running Time]{
   \label{fig:scalability} %% label for first subfigure
    \includegraphics[width=0.22\textwidth]{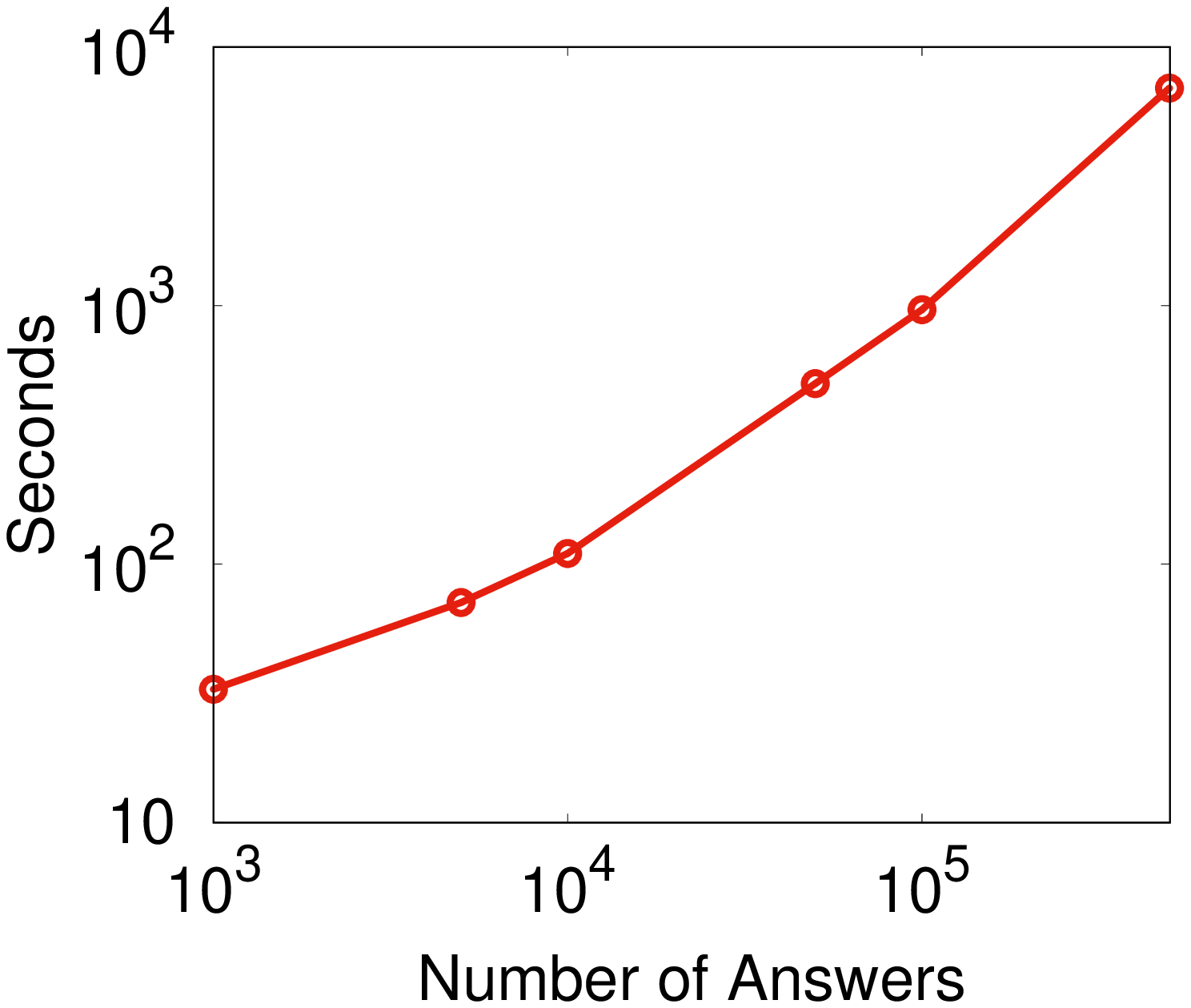}}
  \caption{Efficiency of Truth Inference}
  \label{fig:Efficiency} %% label for entire figure
\end{figure}

In this section, we evaluate the efficiency of T-Crowd. We first investigate the truth inference cost on Celebrity dataset and then show its running time on a single machine. Figure \ref{fig:convergence} shows the change of the objective value in truth inference at each iteration. 
%\nikos{say whether this is average.} \caihua{For one dataset, there is only one convergence curve. Why average? }
Note that our inference model converges to the estimated value, after only a few iterations.

We confirm the low cost of truth inference by measuring the throughput of T-Crowd, i.e., how many answers it can process per second. For this purpose, we use the synthetic data used in Section \ref{sec:exp:synth}, as the number of answers collected for real data are limited. Figure \ref{fig:scalability} shows that the runtime of T-Crowd is approximately linear to the number of answers; T-Crowd can process approximately 100 answers per second on a single machine. This performance is acceptable, given that the rate of incoming answers is much lower in a real crowdsourcing system. It corresponds to the time complexity $\mathcal{O}(w v l  \cdot |\mathcal{A}|)$ at the end of Section \ref{subsec:inference}.
%, indicating the practicality of   T-Crowd, which also includes truth inference as a module of task assignment.
% the objective value decreases fast at the first five iterations and then stay stable. It means that T-Crowd converges quickly in practice.
%$\bullet$ Efficiency of Truth Inference: We use the synthetic data in \ref{sec:exp:synth}. We sample different number of answers to show the running time of T-Crowd on single machine. As shown in Figure \ref{fig:scalability}, the proposed approach has approximately linear complexity in the number of answers. 
%$\bullet$ Efficiency of Task Assignment: 

Finally, we measure the time required to assign a new task to an incoming worker
on the Celebrity dataset. 
We assume that we already obtain the estimated truth from truth inference method.
We show the running time of computing the {\em structure-aware} information gain for all candidate tasks each time a new worker arrives.
% \nikos{confirm}.\caihua{Right}
%We sample the average number of answers per task to show 
Because it is easy to parallelize task assignments, we run eight processes on our machine. 
%\nikos{does this mean that each task assignment costs $\times$8 the time shown in the figure?} \caihua{Yes}
As shown in Figure 
\ref{fig:eff_ass}, the assignment cost increases linearly with the average number of answers collected so far for each task.
This is consistent to our complexity analysis at the end of Section \ref{sec:inherentIG}, which suggests that the cost is linear to the total number $|\mathcal{A}|$ of answers so far.
%It corresponds to the complexity $\mathcal{O}((M+sl) \cdot |\mathcal{A}| )$ we discussed before.  \nikos{++explain}\caihua{Done}
Still, as the figure shows, new assignments can be conducted in real-time, which is important for a real crowdsourcing platform. 
%proposed approach has approximately linear complexity in average number of answers per task. Since the table is fix, it also means it is linear with number of answers.

%!TEX root = main.tex

\section{conclusions}\label{sec:conclusions}

In this paper we design a unified crowdsourcing framework for collecting multi-type tabular data. 
%Heterogeneous structured data brings new unique challenges. 
Most existing methods, which are designed for simple tasks that are all of the same datatype are not effective enough in terms of both truth inference and task assignment. Based on the characteristics of tabular data, we propose a probabilistic truth inference model that unifies worker quality on both categorical and continuous datatypes. Besides, we improve the accuracy of truth inference by considering the variance in the difficulty of different tasks.
In addition, we design an information gain function which we use for selecting the tasks to assign to workers, based on the current answers and the workers' quality. We extend this function to consider the correlation in the quality of certain worker's answers for the same entity.
%These two information gain focus on different characteristic of structured data and both perform well. Using linear combination of these two gains,  we tradeoff them. 
Our experiments on three real datasets and synthetic datasets confirm the superiority of our methods, both in truth inference and task assignment compared to the state-of-the-art. 

In the future, we plan to conduct experiments with larger tables compared to the ones we have used in Section \ref{sec:experiments}. 
%In addition, we plan to extend our approach to apply on tables for which the datatypes or domains of attributes are not known. In this case, the schema of the table should be learnt together with the true values of the entities.  
In addition, we plan to extend our approach to apply on tables for which entities are not known. In this case, entities should also be collected from the crowd.
%\nikos{cite some work on datatype or domain learning}.  \caihua{Changed a little bit.}
A third direction is the acceleration of truth inference and task assignment by parallel and/or distributed computation  as discussed at the end of Section \ref{sec:inherentIG}. Finally, we will explore the possible improvement of our approach by exploiting the possible correlations between entities (not only attributes), e.g., a worker may be more familiar to celebrities starring in a certain category of films or shows.

%$\bullet$ Our experimental datasets described in \ref{sec:experiments} is medium-size, where the number of observations is from 5k to 7k. Evaluations of larger table size are needed. Unfortunately, it can be expensive and time-consuming with much work in crowdsourcing.

%$\bullet$ We assume we know the schema of structured data before collecting, including entities and attributes. But in some times we may not know all entities. It's a good solution to let workers provide and confirm some new entities. Besides, we don't consider the feedback mechanism of workers which allow workers provide more information.

%$\bullet$ Calculation of task assignment is easy to accelerate by parallel or distributed computation. We separate the same number of cells into each machine or process, and return the corresponding information gain. It cost less in communication. As for truth inference, the most time-consuming part is gradient descent. Fortunately, there are many parallel or distributed techniques to accelerate SGD which we can use.

%$\bullet$ We model correlations in the qualities of answers given by the same worker of the same row. However, we can still consider the relationship between entities, like the similarity of the entities, or the domains, which is also useful.

% \balance
\newpage
%{ 
\bibliographystyle{abbrv} \small
\bibliography{refs/paper}
%}

\end{document}